\documentclass{article}
\usepackage[utf8]{inputenc}
\usepackage[english]{babel}
\usepackage{graphicx}
\usepackage{subcaption}
\usepackage{amssymb}
\usepackage{listings}
\usepackage{csquotes}
\usepackage{amsmath}
\usepackage{amsthm}

\usepackage{algpseudocode} 
\usepackage{algorithm}
\usepackage{xcolor}
\usepackage{setspace} 
\usepackage{authblk}

\usepackage[margin=0.9in]{geometry}

\usepackage{comment} 

\title{High Resolution Spatio-Temporal Model for Room-Level Airborne Pandemic Spread}

\author[1]{Teddy Lazebnik\footnote{These authors contributed equally.}\footnote{Corresponding author: t.lazebnik@ucl.ac.uk}}
\author[2]{Ariel Alexi\(^*\)}
\affil[1]{Department of Cancer Biology, Cancer Institute, University College London, London, UK}
\affil[2]{Department of Information Science, Bar-Ilan University, Ramat-Gan, Israel}

\begin{document}

\maketitle

\doublespacing

\begin{abstract}
Airborne pandemics have caused millions of deaths worldwide, large-scale economic losses, and catastrophic sociological shifts in human history. Researchers have developed multiple mathematical models and computational frameworks to investigate and predict the pandemic spread on various levels and scales such as countries, cities, large social events, and even buildings. However, modeling attempts of airborne pandemic dynamics on the smallest scale, a single room, have been mostly neglected. As time indoors increases due to global urbanization processes, more infections occur in shared rooms. In this study, a high-resolution spatio-temporal epidemiological model with air flow dynamics to evaluate airborne pandemic spread is proposed. The model is implemented using a high resolution 3D data obtained using a light detection and ranging (LiDAR) device and computing the model based on the Computational Fluid Dynamics (CFD) model for the air flow and the Susceptible-Exposed-Infected (SEI) model for the epidemiological dynamics. The pandemic spread is evaluated in four types of rooms, showing significant differences even for a short exposure duration. We show that the room's topology and individual distribution in the room define the ability of air ventilation to reduce pandemic spread throughout breathing zone infection.  
\noindent
\textbf{keywords:} Agent-based simulation, Indoor pandemic, Airborne pathogens, Mask-wearing intervention policy, SEI model, Indoor CFD simulation.
\end{abstract}

\section{Introduction}
\label{sec:intro}
Humanity suffered multiple pandemics during its history \cite{pandemic_duration}. In just the last few hundred years, pandemics caused significant mortality, economical crises, and political shifts \cite{pandemic_important}. For example, tens of millions of individuals worldwide died due to the 1918 influenza pandemic \cite{pandemic_important}. Another example is the coronavirus (COVID-19) pandemic which was declared by the World Health Organization (WHO) as a public health emergency of international concern in 2020 and resulted in around six million deaths over the first two years \cite{who_problem,who_data}. 

As such, policymakers are faced with the challenge of controlling a pandemic spread. In particular, this challenge becoming increasingly more relevant, as urbanization grows in the developing world is bringing more people into denser neighborhoods, which results in a higher infection rate, at which new diseases are spread \cite{ravaging1988}. Indeed, Wu et al. (2017) have shown that the overall globalization processes taking place in the recent few decades have facilitated pandemic spread \cite{urbanization2017}. These and other social and economic processes are predicted to make the infectious disease outbreaks nearly constant in the near future \cite{pandemic_duration}. 

Pandemics are taking many shapes such as sexual transmitted diseases (for example, HIV) \cite{std_1,std_2}, social influenced behavior (like, alcoholism) \cite{social_pandemic,social_pandemic_2}, and airborne diseases (such as influenza) \cite{airborne_sample_1,airborne_sample_2}. The group of airborne pandemics are including viruses such as the Lassa virus, Nipah virus or poxviruses, COVID-19, influenza, and others. These are a cause for concern owing to their infection rate and potential for global spread \cite{airborne_important}. Consequently, pandemic intervention policies (PIPs) for airborne pandemics are known to be relatively more harmful to the economy and psychological state of the population relative to other types of pandemic \cite{pip_1,pip_2,pip_3,pip_4,pip_5}. 

Multiple models have been proposed to describe airborne pandemics \cite{models_good_2,models_good_1}, mostly extending the Susceptible-Infected-Recovered (SIR) model proposed by \cite{first_sir}. More often than not, these models are working well for large population sizes and relatively long period of time, providing a fine prediction of the pandemic spread and the influence of a wide range of PIPs on average. However, since these models are focusing on large populations and usually large-scale spatial locations such as cities and countries, they provide less accurate predictions for small size populations in small spatial locations which mainly left neglected.

In this work, we propose a high-resolution spatio-temporal epidemiological model for a case of a single room with a small population. The model is inspired by the \textit{SIR} model and takes into consideration three-dimensional spatial dynamics with air flow. In particular, individuals are infected by breathing pathogen particles from the air and infect others by breathing out pathogen particles. Using the proposed model, one is able to better approximate the airborne pandemic spread in small populations located in a room. The novelty of the proposed model lies in the integration of a computational fluid dynamics simulator for air flow with a spatio-temporal epidemiological model and focusing on a small size population over a short duration. 

The proposed model is evaluated for four types of rooms (classroom, conference room, movie theater, and restaurants), reviling statistically different pandemic spread dynamics. In addition, the influence of the mask-wearing and artificial air ventilation (AAV) PIPs are evaluated for each room type on a wide range of possible configurations. The mask-wearing is shown to better reduce the pandemic spread compared to AAV for all room types. We find that the distribution of individuals in the room has a major influence on the efficiency of AAV, mainly depending on the amount of breathing zone infections.

The paper is organized as follows. Section \ref{sec:related} outlines the current epidemiological and air movement models and their simulations approaches. Section \ref{sec:model} introduces the proposed mathematical model with computer simulation. Section \ref{sec:exp} presents several simulations based on the proposed model. In Sections \ref{sec:discussion} and \ref{sec:conclusion}, we discuss the results and offer future work.

\section{Background}
\label{sec:related}
Multiple studies show that mathematical models and computer simulations are powerful tools for policymakers to investigate pandemic spread and different PIPs with their outcomes in a fast, cheap, and controlled manner  \cite{models_good_1,models_good_2, first_teddy_paper}. There are multiple modeling approaches for epidemiological dynamics \cite{different_approach_from_sir,different_approach_from_sir_2,different_approach_from_sir_3}. The leading approach is extending the \textit{SIR} model \cite{first_sir} with sociological  \cite{models_good_1}, economic \cite{teddy_economic}, biological \cite{models_good_2},
and clinical  \cite{sir_clinical} dynamics to name a few.

More often than not, these models obtain relatively poor results for medium and long prediction periods. One explanation for this shortcoming is the assumption that the population is well mixed which known to be false even for small population sizes and spatial locations \cite{teddy_ariel,graph_2,spatial_1,graph_4,spatial_2,spatial_3}. Indeed, Cooper et al. (2020) used the \textit{SIR} model on the COVID-19 pandemic while relaxing the assumption that the population is mixing homogeneously, showing a fair fitting on six countries with improved results compared to the classical \textit{SIR} model \cite{cooper}. In general, long time periods is hard to predict simply due to statistical fluctuations.

To tackle this challenge, several models introduced spatial dynamics to the spread of a pandemic which can be divided into two main groups: graph-based and metric space. Graph-based models take an abstract approach to modeling the locations in which individuals can be located at. Usually, it is assumed that each node in the spatial graph represents a physical location (such as a room, street, city, or even a country) and that the population is well-mixed in each node \cite{first_teddy_paper,spatial_1}. This approach more often than not ignores the physical properties and dynamics that occur in the location represented by the graph's nodes. For instance, Lazebnik et al. (2021b) proposed a two age-group extended \textit{SIR} model with a three-node graph representing a school, a home, and a work such that the individuals move between them according to their age group and time of the day \cite{teddy_pandemic_management}. Moore and Newman (2000) study several models of disease transmission on small-world networks, in which either the probability of infection by a disease or the probability of its transmission is varied, or both \cite{graph_3}. The authors conducted a numerical analysis which results present a similar behavior to the reported data by \cite{graph_2}. Klovdahl et al. (1994) defined and explored the stochastic \textit{SIR} model on a graph of interactions to represent the pandemic in a cattle trade network with epidemiological and demographic dynamics occurring over the same time scale \cite{graph_4}. The authors used real data on trade-related cattle movements from a densely populated livestock farming region in western France and epidemiological parameters corresponding to an infectious epizootic disease, obtaining fair prediction accuracy. Additionally, Lazebnik and Alexi (2022) proposed a graph-based extended SIR model where each node represents a room in a building \cite{teddy_ariel}. The authors examined the influence of the different moving patterns of the population in several building types (home, school, office, and mall) on the pandemic spread and on the optimal configuration of both spatial and temporal such as mask-wearing and vaccination, respectively. Nevertheless, the authors assumed that the population is well-mixed in each room. As a result, their model produced noisy results for the case of the home-type building, where the population density is low. 

On other hand, for the case of spatial models, it is assumed that the space is continuous and people move in the space over time, representing a more physically accurate representation of movement. For instance, Milner and Zhao (2008) proposed a \textit{SIR} based model where susceptible individuals move away from the previous location of the infection, and all individuals move away from overcrowded regions \cite{spatial_example_1}. Fabricius and Maltz (2020) developed a stochastic \textit{SIR} model with global and local infective contacts  \cite{spatial_example_2}. Paeng and Lee (2017) proposed a \textit{SIR} based model where individuals are assumed to move stochastically within a small fixed radius rather than a random walk \cite{spatial_example_3}. The authors proposed continuous and discrete \textit{SIR} based models that show spatial distributions. They show that the propagation speed and size of an epidemic depend on the population density and the infectious radius.

In the context of airborne pandemics, one can focus on the infection that occurs in a room by tracking the air flow with the pathogen particles it carries between the individuals in the room \cite{airflow_pandemic}. Wei and Li (2016) review the release, transport, and exposure of expiratory droplets because of respiratory activities in the context of a pandemic indoor \cite{indoor_pandemic}. The authors concluded that droplets or droplet nuclei are transported by air flow, which is sometimes affected by the human body plume. They suggested that the usage of a face mask, as well as room air ventilation, can reduce the infection rate. 

Air flow (or air movement) dynamics are vastly explored in multiple contexts such as healthcare \cite{airflow_health}, mechanics \cite{airflow_mechanics}, agriculture \cite{airflow_agriclature}, and epidemiology \cite{airflow_pandemic}. In particular, Peng et al. (2020) explored the pandemic spread of the airborne COVID-19 pathogen in indoor settings using a combination of the box and Wells-Riley models \cite{indoor_pandemic_model,wells_riley}. The authors have derived an expression for the number of secondary
infections. Nonetheless, their model is not taking into consideration a detailed, and accurate representation of the room's geometry. Moreover, the box model used by the authors does not correctly represent common cases such as rooms where clear directional flow or infection due to overlapping breathing zones. In this work, we aim to tackle these challenges using more detailed air flow dynamics.

There are various modeling approaches to predict the air flow within buildings such as Multi-zone models and Zonal models. However, Computational Fluid Dynamics (CFD) models considered to be the most accurate for a single room \cite{airflow_building_review,airflow_building_review_2}.

The CFD models are numerical methods of solving fluid flow using the Navier-Stokes (NS) equations \cite{ns_3d}. These models use numerical algorithms to integrate the NS equations over a given mesh by converting the integral equations to algebraic equations (e.g., discretization) and then solving them iteratively \cite{cfd_intro}. In the context of building air flow or even room-level air flow, the CFD modeling approach subdivides an individual room into many space segments and each one is treated as an atomic segment in which the NS equations are computed \cite{airflow_building_review}. There are three main types of CFD implementation for indoor air flow dynamics: Reynolds-Averaged Navier-Stokes (RANS), Large Eddy Simulation (LES), and Direct Numerical Simulation (DNS) \cite{airflow_building_models}.

The RANS equations take advantage of the Reynolds decomposition technique whereby an instantaneous quantity is decomposed into fluctuating quantities and the average value of some time duration. This technique provides approximations over time of the NS equations, averaging the results on short periods \cite{rans}. Complementary, DNS numerically solves full Navier–Stokes equations using a very fine mesh to capture all the scales that are present in a given flow. Interrelate, LES computes large-scale motions similarly to DNS but with a larger grid, and sub-grid scale dynamics are solved using an averaging method such as the RANS approach. While all three CFD computational approaches are able to provide accurate predictions for air flow, the RANS approach is the most popular one for the indoor environment. This is because the other two CFD approaches are significantly more computationally expensive without a justified improvement in prediction accuracy for most cases \cite{cfd_summary}. 

Several CFD-based models and simulators have been developed for indoor context \cite{cfd_1,cfd_2,cfd_3}. Hiyama and Kato \cite{cfd_1} developed a 3D CFD model for close space with air flow in whole building settings. The authors integrated the outcomes of the CFD simulation with building energy simulations to achieve a more accurate time-series analysis of building energy consumption compared to conventional energy simulations \cite{cfd_1}. Nahor et al. \cite{cfd_2} proposed a 3D CFD model to calculate the velocity, temperature, and moisture distribution in an existing empty and loaded cool store. The authors validated their model with data from several experiments and show that the model was capable of predicting both the air and product temperature with
reasonable accuracy \cite{cfd_2}. Smale et al. \cite{cfd_3} reviewed the application of CFD and other numerical modeling techniques to the prediction of air flow in refrigerated food applications including cool stores, transport equipment, and retail display cabinets. The authors show that given enough computation power, CFD-based models obtain high accuracy in all these tasks. For the case of an individual breathing in a room, Cravero and Marsano shown that the CFD model provides a fine accuracy to the air movement including breathing dynamics, air ventilation, and air diffusion \cite{cfd_cool}. Thusly, CFD-based models are accurately describing the air movement dynamics across a wide range of environments in general and room types in particular. As such, the CFD model implemented using the RANS method is used in this research.

\section{Model Definition}
\label{sec:model}
\begin{figure}[!ht]
    \centering
    \includegraphics[width=0.4\textwidth]{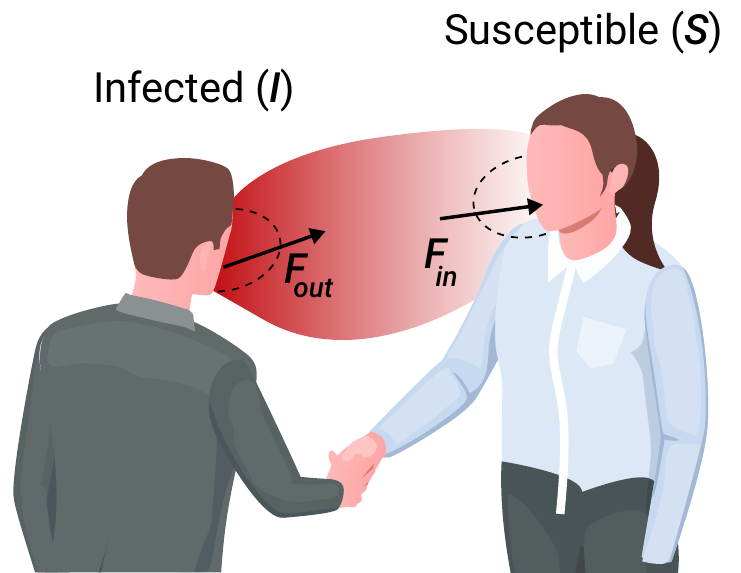}
    \caption{A schematic view of the model's dynamics. \(F_{out}\) and \(F_{in}\) are the force of air getting out and in (with the pathogen's particles) during an individual's breathing process.}
    \label{fig:model_view}
\end{figure}

The proposed model is constructed from three main components: a population of individuals, an environment (room) representing a three-dimensional space, in which the population is located, and air flow dynamics taking place. A schematic view of the model's dynamics is shown in Fig. \ref{fig:model_view} for the simplistic case of two individuals interacting. During exposure to pathogens particles in the air by the infected individual, the second individual has been infected by breathing in the pathogen particles. Formally, the model is defined by a tuple \(M := (\mathbb{P}, E)\) where \(\mathbb{P}\) is a set (the population) of individuals and \(E\) is the environment of the model representing the room and air flow. The components of the tuple are described below in detail.

\subsection{The environment \(E\)}
The spatial component of the model is considering a three-dimensional continuous space \(E \subset \mathbb{R}^3\) of length, width, and height of \(L, W,\) and \(H\), respectively. The space \(E\) is discretized in all three dimensions with size \(\delta\), represented by a matrix \(M_E \in \mathbb{R}^{L/\delta \times W/\delta \times H/\delta}\). Each cell \((i, j, k)\) of the space matrix \(M_E\) representing a cube with a center of mass at \(((i+0.5) \cdot \delta, (j+0.5) \cdot \delta, (k+0.5) \cdot \delta)\) and with a volume \(\delta^3\). 

The space \(E\) contains a set of objects and individuals. Formally, each location in the space \((i, j, k) \in E\) is either filled by an object (which can be an individual) or by air. If a location is fulfilled by an object, the air is not presented, and vice versa. We assume that all objects in the space are static and therefore do not move over time. A location with air has four properties: pathogen particles \(\lambda \in \mathbb{N}\), velocity \(v \in \mathbb{R}^3\), acceleration \(a \in \mathbb{R}^3\), and pressure \(\rho \in \mathbb{R}^3\). Furthermore, we assume that the air in the space is incompressible.  

The air in the space is moving based on the incompressible Navier-Stokes equations \cite{ns}:
\begin{equation}
    \frac{\partial \vartheta}{\partial t} + u \cdot \nabla \vartheta = \frac{- \nabla \rho}{\rho} + \mu \nabla ^2 u + g,
    \label{eq:ns}
\end{equation}
where \(\vartheta = (u, v, w)^T\) is the velocity vector, \(t\) is time, \(\rho\) is the uniform density of the atmosphere, \(\mu\) is the dynamic viscosity, and \(g\) is the gravity constant vector. The pathogen particles are airborne and move with the air as a result. In addition, as pathogen particles can not survive outside the host for too long, the number of the pathogen particles generated by an infected individual in a single breath decay exponentially over time \cite{particale_decay,particale_decay_2}. As such, for each \((i, j, k) \in E\), the number of pathogen particles satisfies:
\begin{equation}
    \frac{\partial C}{\partial t} = C \big ( \nabla \tau - \nabla u - D \big ),
    \label{eq:particles_in_room}
\end{equation}
where \(C(x, t)\) is the number of pathogen particles in location \(x\) at time \(t\), \(\tau\) is the tensor of turbulent diffusivity, and \(D\) is the decay rate of the pathogen particle.

For the initial conditions, it is assumed that the entire air is without pathogen particles and \(\forall x: u(x, 0) = 0\) (i.e., the “static” state). The boundary conditions are defined to be the boundary cells of the space matrix \(M_E\) and all locations that contain an object. Namely, open boundary conditions (stress-free boundary conditions) are specified at the outlet. No-slip boundary conditions are provided in all four boundary directions \cite{particale_decay}.

\subsection{The population \(\mathbb{P}\)}
\label{section:population}
The model considers a constant population \(\mathbb{P}\) with a fixed number of individuals \(N \equiv |\mathbb{P}|\). Each individual belongs to one of three groups: susceptible \((S\)), exposed \((E\)), or infected \((I\)), such that \(N = S + E + I\). Individuals in the susceptible group have no immunity and are susceptible to infection. When an individual in the susceptible group (\(S\)) is exposed to the pathogen, the individual is transferred to the exposed group (\(E\)) at a rate \(\beta\). The individuals stay in the exposed group \((E)\) for \(\xi\) time steps. This duration is mainly defined by the pathogen and its interaction with the immune system. After \(\xi\) time steps, the individuals are transferred to the infected group \((I)\). 

This \textit{SEI} model is a private case of the \textit{SEIR} model such that \(\forall t: R(t) = 0\). This case better suits the short-duration pandemic spread at the beginning of the pandemic, where there is still a small (or even neglected) number of recovered or vaccinated individuals, such as the one of interest in this model.

Formally, let \(\mathbb{P}\) be a non-empty set of individuals such that each individual \(p \in \mathbb{P}\) is defined by a tuple \( p :=  (l, b, \nu, c)\), where \(l = (l_x, l_y, l_z, l_\alpha, l_\beta)\) is the location and orientation of the individual's face such that \((l_x, l_y, l_z) \in E\) is the three-dimensional location in the environment and \(l_\alpha \in [0, 2 \pi], l_\beta \in [0, \pi]\) are the angles in the \(xy\) and \(xz\) plans, respectively; \(b = (\phi_\delta, \phi_r, \phi_g)\) are the breathing-related properties such that \(\phi_\delta\) is the minimal number of pathogen particles needed to be inside the individual's body to cause infection and making an susceptible individual to become exposed to the pathogen, \(\phi_r\) is the natural reduction rate of the pathogen particle for susceptible individuals, \(\phi_g\) is the amount of pathogen particle the individual generates during breathing out when it is infected; \(\nu \in \{s, e, i\}\) is the individual's current epidemiological state; and \(c\) is the number of pathogen particles currently found at the individual's body. 

At each point in time, each individual in the population is either breathing in, out, or neither. During the inhaling phase, a force \(F_{in}\) originated at \((l_x, l_y, l_z)\) is manifested. The \(F_{in}\) force influences the air movement as well as the pathogen particles in its close proximity. In particular, the pathogen particles that are located at a distance \(\gamma\) from \((l_x, l_y, l_z)\) during the inhaling are removed from the environment and inserted into the individual's body. In a similar manner, during an exhaling phase, a force \(F_{out}\) originated at \((l_x, l_y, l_z)\) is manifested. If the individual is infected (i.e., \(\nu = i\)) than \(\phi_g\) pathogen particles are generated at each breath and added to the environment in a exponentially decaying distribution as a function of the distance from \((l_x, l_y, l_z)\) with weights \(\omega_x, \omega_y,\) and \(\omega_z\) for the \(x, y,\) and \(z\) axis. Individuals follow a breathing pattern as follows: inhaling, not breathing, exhaling, and not breathing. This pattern is repeated for each individual during the entire model's dynamics.

An susceptible individual (\(\nu = s\)) becomes exposed (\(\nu = e\)) if \(c > \phi_\delta\). During the time which an individual is belongs to either the exposed or infected epidemiological states, the change in the number of pathogen particles in the body \((c)\) does not have any effect on its epidemiological state. 

The population component (\(\mathbb{P}\)) is implemented using the agent-based simulation method to allow heterogeneous population  \cite{agent_based_with_ai,agent_based_exp_1,agent_simulation_1} where each individual in the population has a unique values of the inhaling and exhaling duration and volume as well the number of pathogen particles needed to make it infected. For a more detailed implementation description see \cite{teddy_pandemic_management}.

\subsection{Model implementation}
\begin{figure}[!ht]
    \centering
    \includegraphics[width=0.99\textwidth]{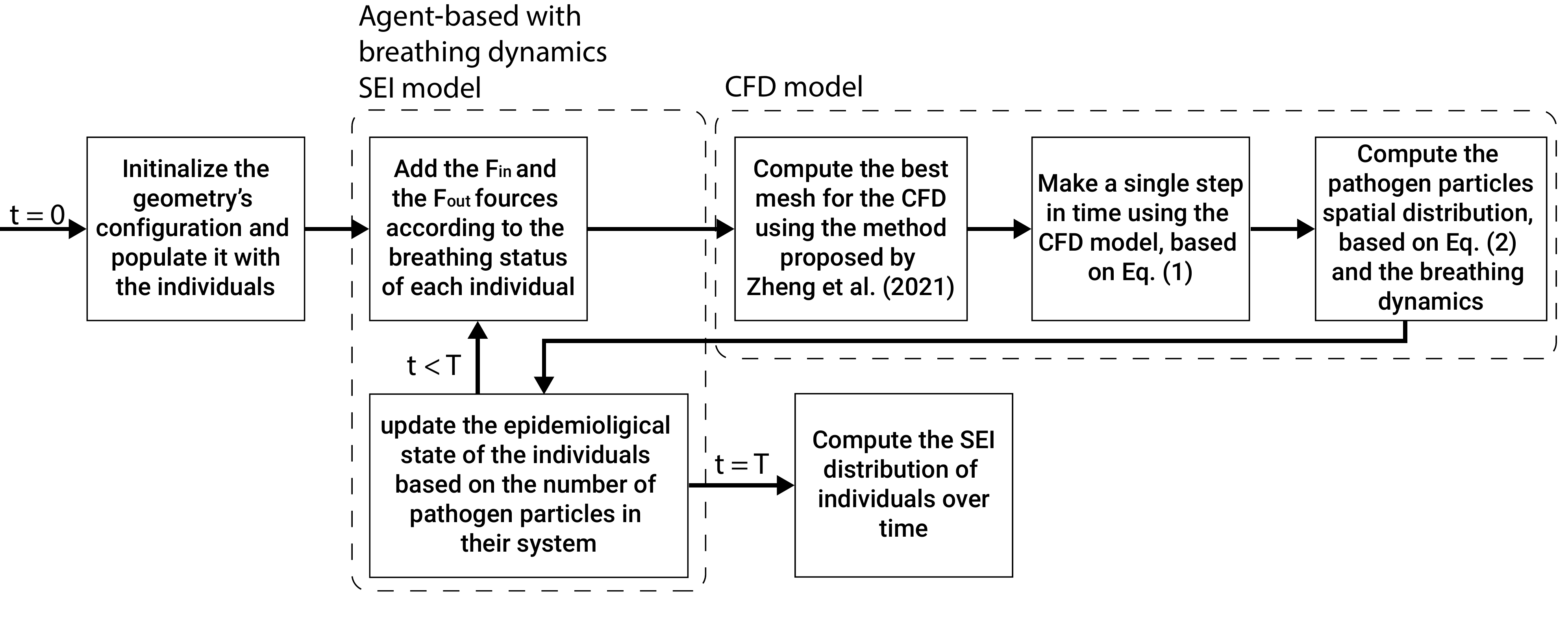}
    \caption{A schematic view of the simulator's process, including the integration between the agent-based with breathing dynamic.}
    \label{fig:sim_view}
\end{figure}

The proposed model is implemented as a computer simulation. The environment \(E\) is represented by a 3D tensor such that each node represents a space segment with a volume of \(\delta^3\). At the beginning of the simulation (i.e., \(t = 0\)), the individuals that occupy the environment are set in the environment such that each individual is allocated with a unique inhaling and exhaling duration and volume values from a pre-defined distribution. In a similar manner, a number of pathogen particles needed to make an individual infected are randomly sampled from a given distribution and set to each individual. Afterward, at each step in time, five computational steps take place. First, based on the individuals' stage in the breathing cycle (inhaling, exhaling, neither) a force \((F_{in}\), \(F_{out}\), none) is allocated to the corresponding location of the individuals' face, respectively (see Section \ref{section:population}). Moreover, if an infected individual is exhaling, \(\phi_g\) pathogen particles are introduced to the room at \((l_x, l_y, l_z)\). Second, an adaptive mesh for the CFD model is computed using the method proposed by \cite{particale_decay} such that at least a single node of the mesh is located in each node in the \(M_E\) matrix in order to ensure a lower boundary of the size of the geometry's resolution. Third, the air movement in a single step in time is computed using the RANS method \cite{rans} for the CFD model (practically approximating Eq. (\ref{eq:ns}) for the mesh's nodes' location in a specific point in time). Fourth, Using the \(\nabla u\) value computed on the geometry's mesh from the previous step, Eq. (\ref{eq:particles_in_room}) is computed and the number of pathogen particles that enter individuals' bodies are updated. Finally, the epidemiological state of each individual in the population is updated based on the number of pathogen particles in its system and its current epidemiological state. This process repeats itself for \(T\) time steps.  A schematic view of the simulator's process is presented in Fig. \ref{fig:sim_view}.

\section{Simulations and Results}
\label{sec:exp} 
\begin{figure}[!ht]
    \centering
    \includegraphics[width=0.99\textwidth]{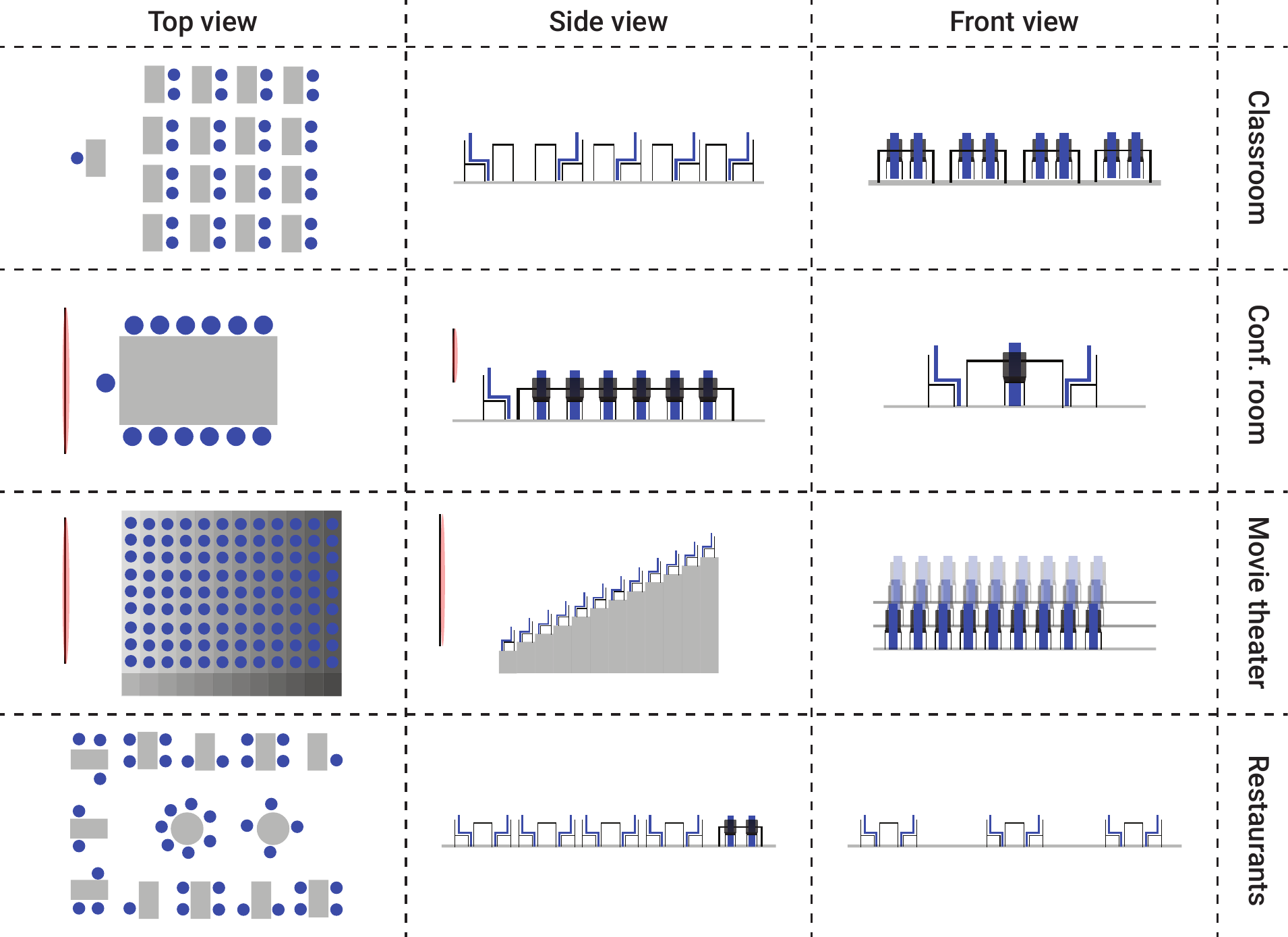}
    \caption{Schematic 3D projection views of the four room types.}
    \label{fig:setup}
\end{figure}

We evaluated the model on four types of rooms: classrooms, conference rooms, movie theaters, and restaurants. We picked these types of rooms due to three reasons: First, most of the individuals in the room are not moving most of the time. We tried to avoid movement as it influences the pandemic spread significantly \cite{indoor_pandemic,teddy_ariel,human_behavior_in_buildings} and not in the scope of the proposed model. Second, the room types should have a unique and distinguished topology. Third, the rooms should involve at least several individuals in order to allow a pandemic spread signal to be noticeable. Since classrooms, conference rooms, movie theaters, and restaurants fulfill these conditions \cite{restaurant,restaurant_2}, we used them. Moreover, each one of the room types has a unique topology and distribution of the population inside the room. These differences are shown in a schematic 2D projection of each room type into a top, side, and front view of the 3D structure of the room in Fig. \ref{fig:setup}. Other researchers also investigated pandemic spread inside buses \cite{other_buildings_1}, workshop rooms \cite{other_buildings_1}, homes \cite{teddy_ariel}, and gyms \cite{other_buildings_2}. However, these do not satisfy at least one of our requirements as individuals are moving quite significantly in buses, workshop rooms, and gyms. The population size in homes is usually too small to efficiently evaluate pandemic spread. 

The rooms' topology were obtained using a 3D LiDAR-based scanning\footnote{Using the technology available in the iPhone 12 device with the \textit{RoomScan LiDAR} application.} of the room and mapped it into its 3D geometry \cite{lidar}. In particular, the LiDAR program captures a 3D cloud point which is later processed to a mesh \cite{point_to_mesh}. The obtained mesh's resolution is one cubic centimeter. This mesh is used as the boundary condition of the geometry. For each room, we obtained the places where individuals located (sit) based on the chairs' locations and added \(0.8\) meters on the z-axis for their faces' location. Each individual occupy a box in space with \(170_{cm}\) height, \(55_{cm}\) width, and \(40_{cm}\) length. In addition, each individual starts the breathing cycle in a random time between \(t = 0 \) and \(t = F_{in}^T + F_{out}^T\). 

Moreover, for the epidemiological properties, we simulated the COVID-19 pandemic with data from March, 1st 2020 to September, 1st 2020 in Israel, obtained from the WHO \cite{who_data}. This date range is chosen since, except for a single lockdown, no significant PIPs were used, no vaccination is issued and only one strain dominated the population. Thus, this period in Israel best suits the \textit{SEI} dynamics and as such yields as realistic epidemiological value (e.g., exposure to infection duration) as available. 

We assume only one individual (e.g., the index individual) is infected at the beginning of each simulation. As such, the other individuals are susceptible. Formally, the initial condition of the simulator is assumed to be: \( S(0) = N - 1, \; E(0) = 0, \; I(0) = 1\). Namely, our initial conditions consider only a single infected individual which is taken randomly from one realisation to another.

A summary of the used model parameters' values is provided in Table \ref{table:parameters}. In addition, a summary of the room types with a qualitative description of the population and the size of the room is provided in Table \ref{table:topologies}.

\begin{table}[!ht]
\centering
\begin{tabular}{|p{0.7\textwidth}|p{0.1\textwidth}|p{0.1\textwidth}|p{0.1\textwidth}|}
\hline
\textbf{Parameter} & \textbf{Notation} & \textbf{Value} & \textbf{Source} \\ \hline
Population size [\(1\)] & \(N\) & DPS & - \\ \hline
Exposed to infected transformation rate in minutes [\(\frac{1}{t}\)] & \(\xi\) & \(1.01 \cdot 10^{-4}\) & \cite{teddy_ariel} \\ \hline
Simulation's step in time (in seconds) [\(t\)] & \(\Delta t\) & 0.02 & \cite{airflow_building_review} \\ \hline
Number of simulation steps [\(1\)] & T & 270,000 & - \\ \hline
Average decay rate of the pathogen particles in air in minutes [\(\frac{1}{t}\)] & \(D\) & \(5.5 \cdot 10^{-3}\) & \cite{particale_decay} \\ \hline
CFD's mesh's single volume element size in cubic centimeters [\(m^3\)] & \(\delta\) & \(1\) & \cite{particale_decay_3} \\ \hline
Inhaling duration in seconds [\(t\)] & \(F_{in}^{\tau}\) &  \(1.42 \pm 0.25\) &  \cite{breathing_time,breathing_volume} \\ \hline
Exhaling  duration in seconds [\(t\)] & \(F_{out}^{\tau}\) &  \(2.28 \pm 0.47\) &  \cite{breathing_time_2,breathing_volume} \\ \hline
No breathing duration in seconds [\(t\)] & - & \(0.39 \pm 0.04\) &  \cite{breathing_volume,breathing} \\ \hline
Inhaling volume in cubic centimeter [\(m^3\)] & \(|F_{in}| / F_{in}^{\tau}\) & \(304 \pm 71\)  &  \cite{breathing_volume} \\ \hline
Exhaling volume in cubic centimeter [\(m^3\)] & \(|F_{out}| / F_{out}^{\tau}\) &  \(198 \pm 41\) &  \cite{breathing_volume} \\ \hline
Average distance of pathogen particle influenced by inhaling in meters [\(m\)] & \(\gamma\) & 0.34 & \cite{bio_data} \\ \hline
Average decay rate of pathogen particles in host in minutes [\(\frac{1}{t}\)] & \(\phi_r\) & \(6.6 \cdot 10^{-4}\) & \cite{bio_data} \\ \hline
Average number of pathogens particles needed to infected a susceptible individual [\(1\)] & \(\phi_\delta\) & \(10^8\) &  \cite{covid_talbe_data} \\ \hline
Average number of pathogens particles generated by infected individual at each exhaling [\(1\)] & \(\phi_g\) & \(1.3 \cdot 10^{7}\) & \cite{covid_talbe_data} \\ \hline
\end{tabular}
\caption{The model's parameters with their default values. DPS stands for differing per simulation which means the value is dependent on the specific instance of the simulation computed. }
\label{table:parameters}
\end{table}

\begin{table}[!ht]
\centering
\begin{tabular}{|p{0.15\linewidth}|p{0.15\linewidth}|p{0.3\linewidth}|p{0.2\linewidth}|}
\hline
\textbf{Name} & \textbf{Population size} & \textbf{Room size [meter]} & \textbf{Density [\(1/meter^3\)]} \\ \hline
Classroom 1 & \(N = 33\) & \(L=10.00, W=6.00, H=2.60 \) & \(0.211\) \\ \hline
Classroom 2 & \(N = 28\) & \(L=8.00, W=5.63, H=2.88 \) & \(0.215\)  \\ \hline
Classroom 3 & \(N = 27\) & \(L=7.80, W=7.70, H=2.61 \) & \(0.173\)  \\ \hline
Classroom 4 & \(N = 31\) & \(L=9.00, W=7.00, H=3.04 \)  & \(0.164\) \\ \hline
Classroom 5 & \(N = 36\) & \(L=8.77 , W=14.25, H=3.20 \)  & \(0.090\) \\ \hline 

Restaurant 1 & \(N = 19\) & \(L=12.12 , W =8.25 , H=2.75  \) & \(0.069\)  \\ \hline 
Restaurant 2 & \(N = 23\) & \(L=25.00 , W=14.00 , H=2.83  \) & \(0.023\)  \\ \hline 
Restaurant 3 & \(N = 37\) & \(L=26.25 , W=20.00 , H=3.63  \)& \(0.019\)  \\ \hline 
Restaurant 4 &  \(N = 48\) & \(L=21.88 , W=16.00 , H=3.00  \)& \(0.046\)  \\ \hline 
Restaurant 5 &  \(N = 67\) & \(L=13.87 , W=10.95, H=4.40  \) & \(0.100\) \\ \hline 

Movie theater 1 & \(N = 240\) & \(L=22.89, W=18.00, H=11.25 \) & \(0.052\) \\ \hline 
Movie theater 2 & \(N = 112\) & \(L=12.72, W=10.00, H =7.00 \)& \(0.126\)  \\ \hline 
Movie theater 3 &  \(N = 270\) & \(L=24.72 , W=19.44 , H=12.15 \) & \(0.046\) \\ \hline 
Movie theater 4 & \(N = 400\) & \(L=18.94, W = 14.80, H =9.30 \) & \(0.153\) \\ \hline 
Movie theater 5 & \(N = 720\) & \(L=36.00, W=28.30, H=17.68 \)& \(0.040\) \\ \hline 

Conference 1 & \(N = 9\) & \(L=6.50, W=4.00, H=3.12  \)& \(0.110\)  \\ \hline
Conference 2 &  \(N = 11\) & \(L=7.98, W=3.76 , H=2.79 \)& \(0.131\)  \\ \hline
Conference 3 & \(N = 16\) & \(L=10.00, W=4.09, H=2.90  \) & \(0.135\) \\ \hline
Conference 4 & \(N = 17\) & \(L=11.84, W=5.20, H=2.91  \)& \(0.095\)  \\ \hline
Conference 5 & \(N = 24\) & \(L=11.21, W=6.70, H=2.81 \)& \(0.114\)  \\ \hline

\end{tabular}

\caption{The different room types used in the experiments with a qualitative description of the population size and the room's dimensions in meters.}
\label{table:topologies}
\end{table}

\subsection{Baseline model dynamics}
\begin{figure}
\centering
\begin{subfigure}{.49\linewidth}
    \centering
    \includegraphics[width=0.99\linewidth]{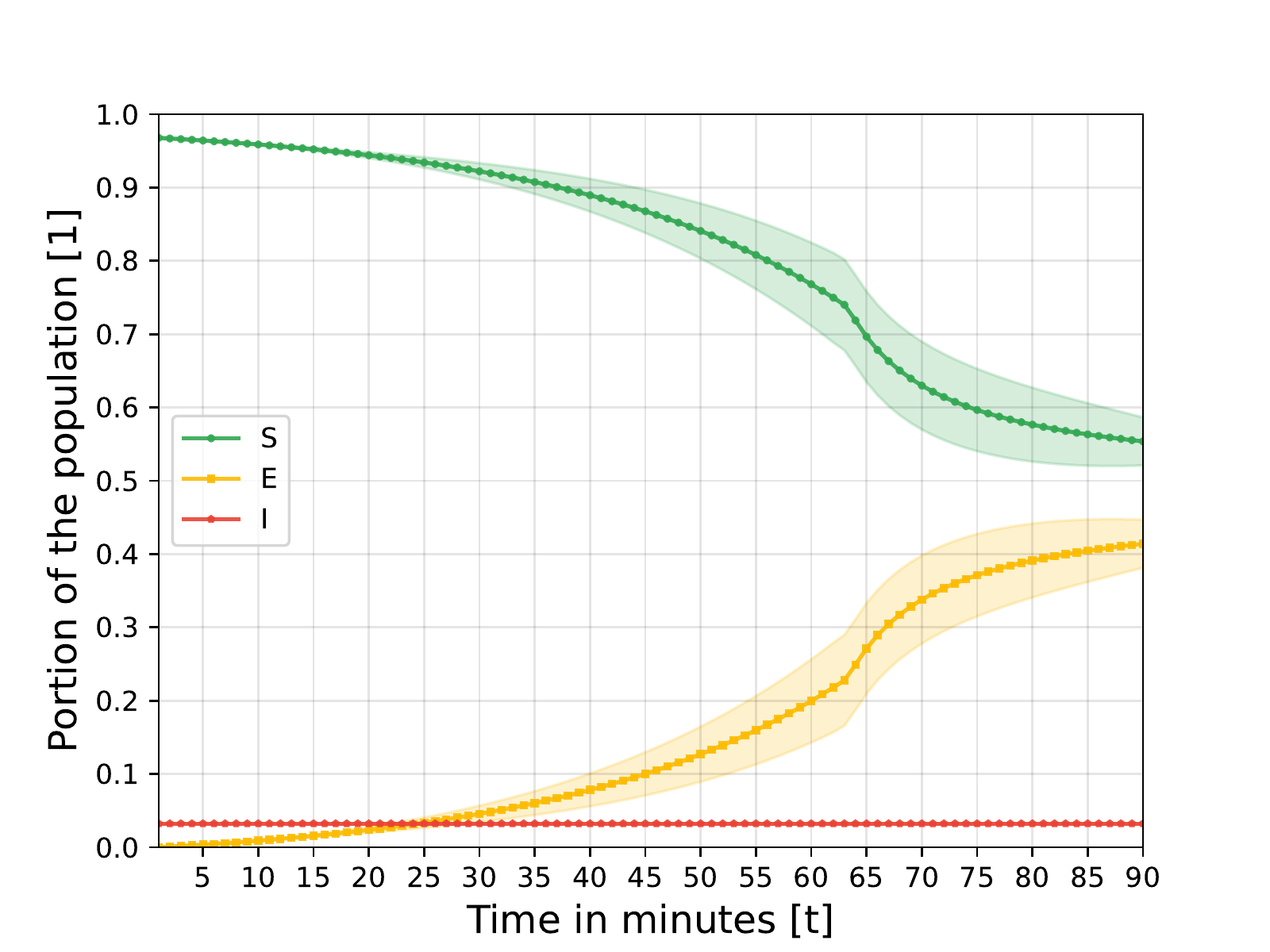}
    \caption{Classroom.}
    \label{fig:baseline_classroom}
\end{subfigure}
\begin{subfigure}{.49\linewidth}
    \centering
    \includegraphics[width=0.99\linewidth]{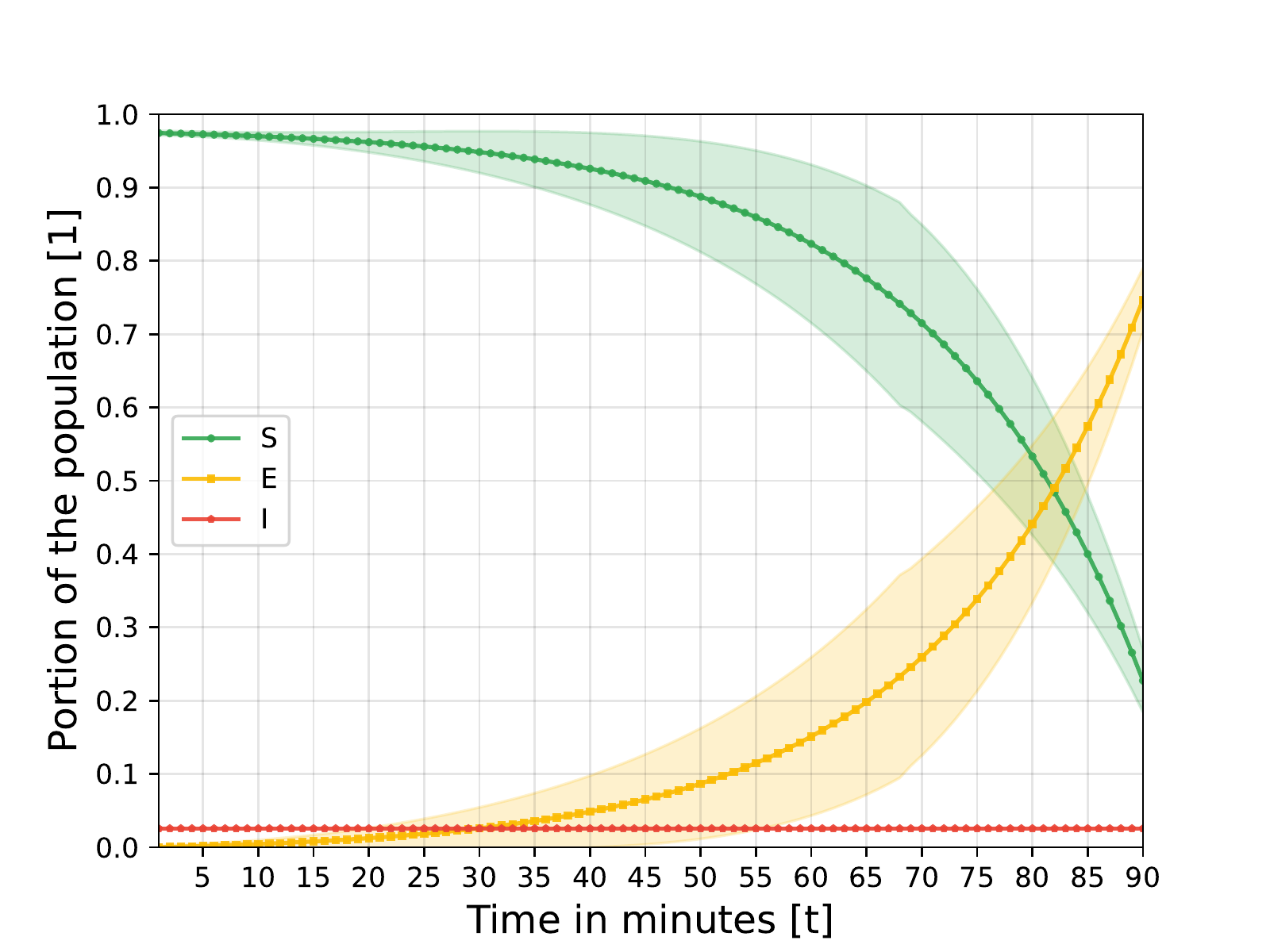}
    \caption{Conference room.}
    \label{fig:baseline_conf}
\end{subfigure}

\begin{subfigure}{.49\linewidth}
    \centering
    \includegraphics[width=0.99\linewidth]{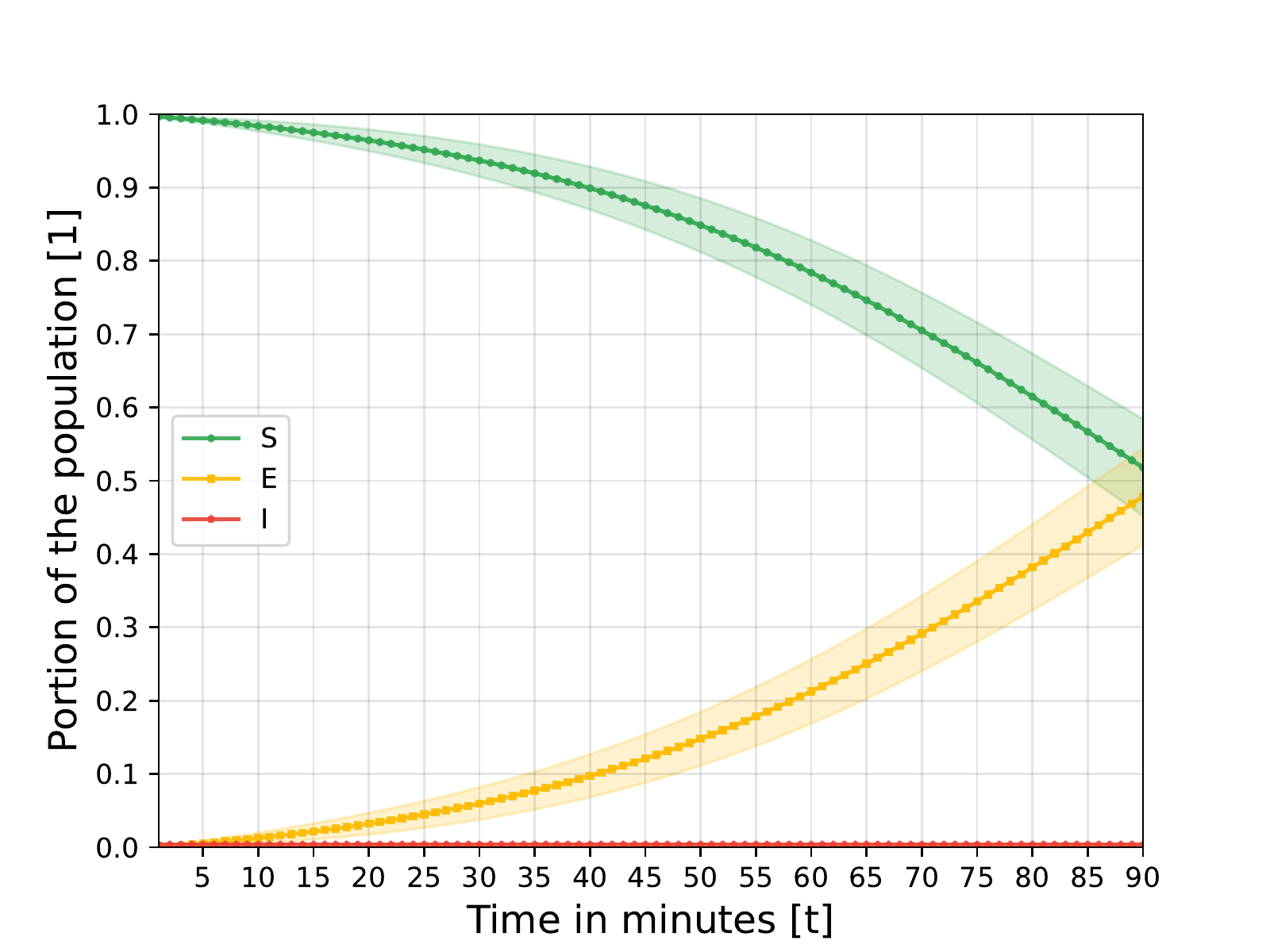}
    \caption{Movie theater.}
    \label{fig:baseline_movie}
\end{subfigure}
\begin{subfigure}{.49\linewidth}
    \centering
    \includegraphics[width=0.99\linewidth]{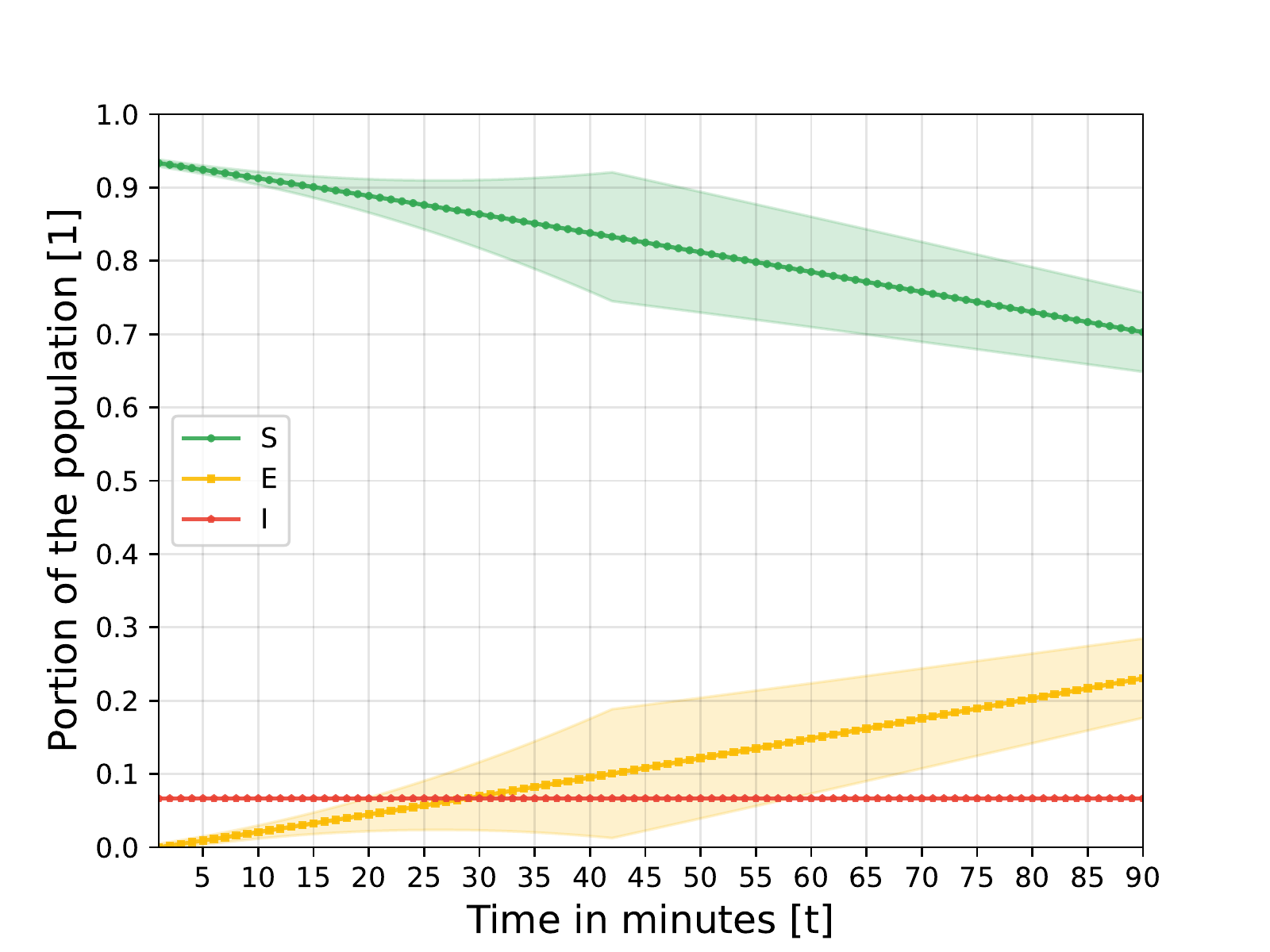}
    \caption{Restaurant.}
    \label{fig:baseline_restaurant}
\end{subfigure}
\caption{The \textit{SEI} over time for each type of room, shown as mean \(\pm\) standard deviation. The results are computed as a result of 100 realisations for each room and five rooms for each room-type (\(n = 500\)).}
\label{fig:baseline}
\end{figure}

In order to evaluate the airborne pandemic spread dynamics in the different types of rooms, we computed for each room the pandemic spread for 90 minutes. Each time, picking a different infected individual in a uniform distribution. We repeated the simulation 100 times. The results of the simulation are shown in Fig. \ref{fig:baseline}, where the x-axis is the time passed in minutes from the beginning of the pandemic and the y-axis indicates the portion of the population in each epidemiological state where \(S\), \(E\), \(I\), and \(R\) stands for susceptible, exposed, infected, and recovered, respectively. 

\subsection{Pandemic intervention policies}
PIPs are all the actions of individuals that aim to control the pandemic spread. Indeed, policymakers uses multiple types of PIPS such as lockdown \cite{lockdown_sir} and artificial job separation \cite{teddy_economic}. For airborne pandemics, the mask-wearing PIP is considered efficient and used for the COVID-19 pandemic \cite{masks_good} and the influenza pandemic \cite{influenza_mask} to name a few. However, in practice, the masks differ in their effectiveness and the portion of the population that wears them \cite{close_lockdown,masks_n95,masks_good}. Hence, we define the mask-wearing PIP using two parameters \(\eta\) and \(\kappa\) that stand for the portion of the population that wearing a mask and the portion of the pathogen particles filtered by the mask. The reduction of the mask's effectiveness over time is neglected due to the short-horizon duration of the simulation \cite{masks_n95}. 

In addition, air ventilation is also considered an effective PIP for airborne pandemics \cite{indoor_pandemic,indoor_pandemic_model}. Unlike other PIPs, the air ventilation is highly affected by the room's topology in directed and undirected ways. For example, the location of a window, alongside its size and the outdoor air temperature can alter the pandemic spread. Thus, we defined a simplified version of air ventilation. Assuming a centralized air ventilation system that removes a portion \(\chi\) of the pathogen particles from the air in the room in a uniform way and occurs in a discreet manner every \(\zeta\) time steps. This definition is identical to mixing the air in the room with fresh air at a proportional portion in a uniformly distributed way.

\begin{figure}
\centering
\begin{subfigure}{.4\linewidth}
    \centering
    \includegraphics[width=0.75\linewidth]{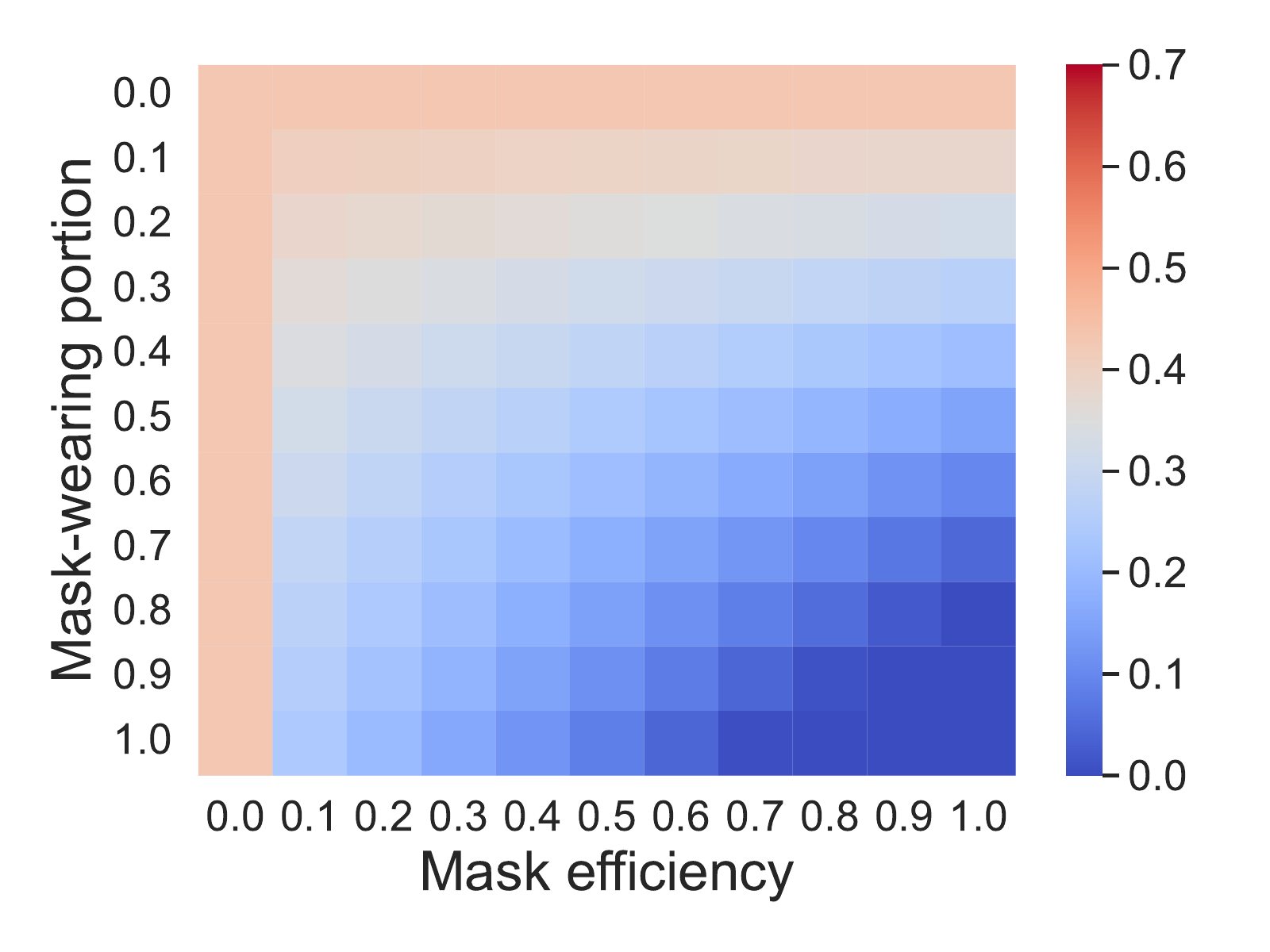}
    \caption{Classroom - masks.}
    \label{fig:pip_classroom_masks}
\end{subfigure}
\begin{subfigure}{.4\linewidth}
    \centering
    \includegraphics[width=0.75\linewidth]{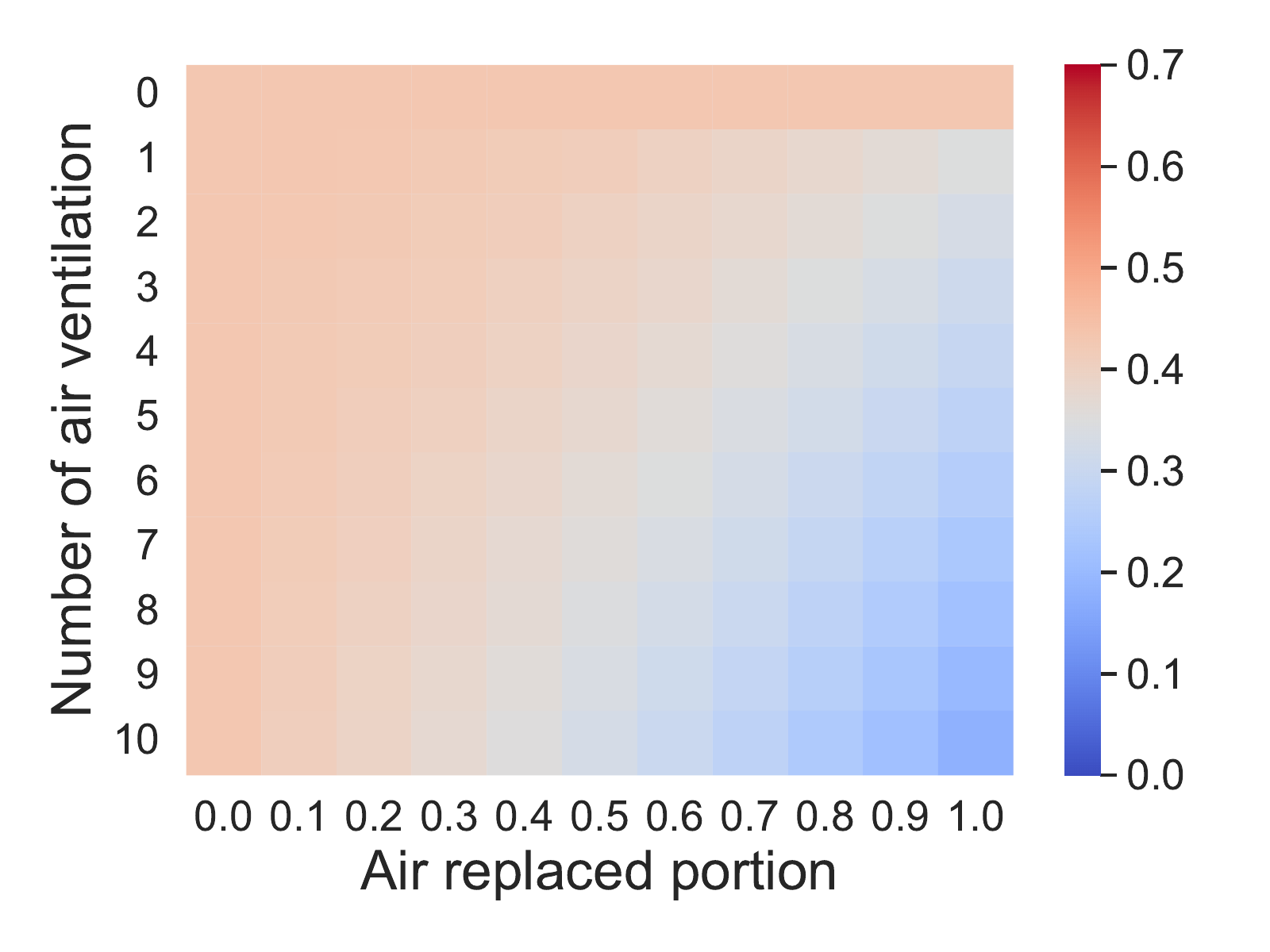}
    \caption{Classroom - air ventilation.}
    \label{fig:pip_classroom_air}
\end{subfigure}

\begin{subfigure}{.4\linewidth}
    \centering
    \includegraphics[width=0.75\linewidth]{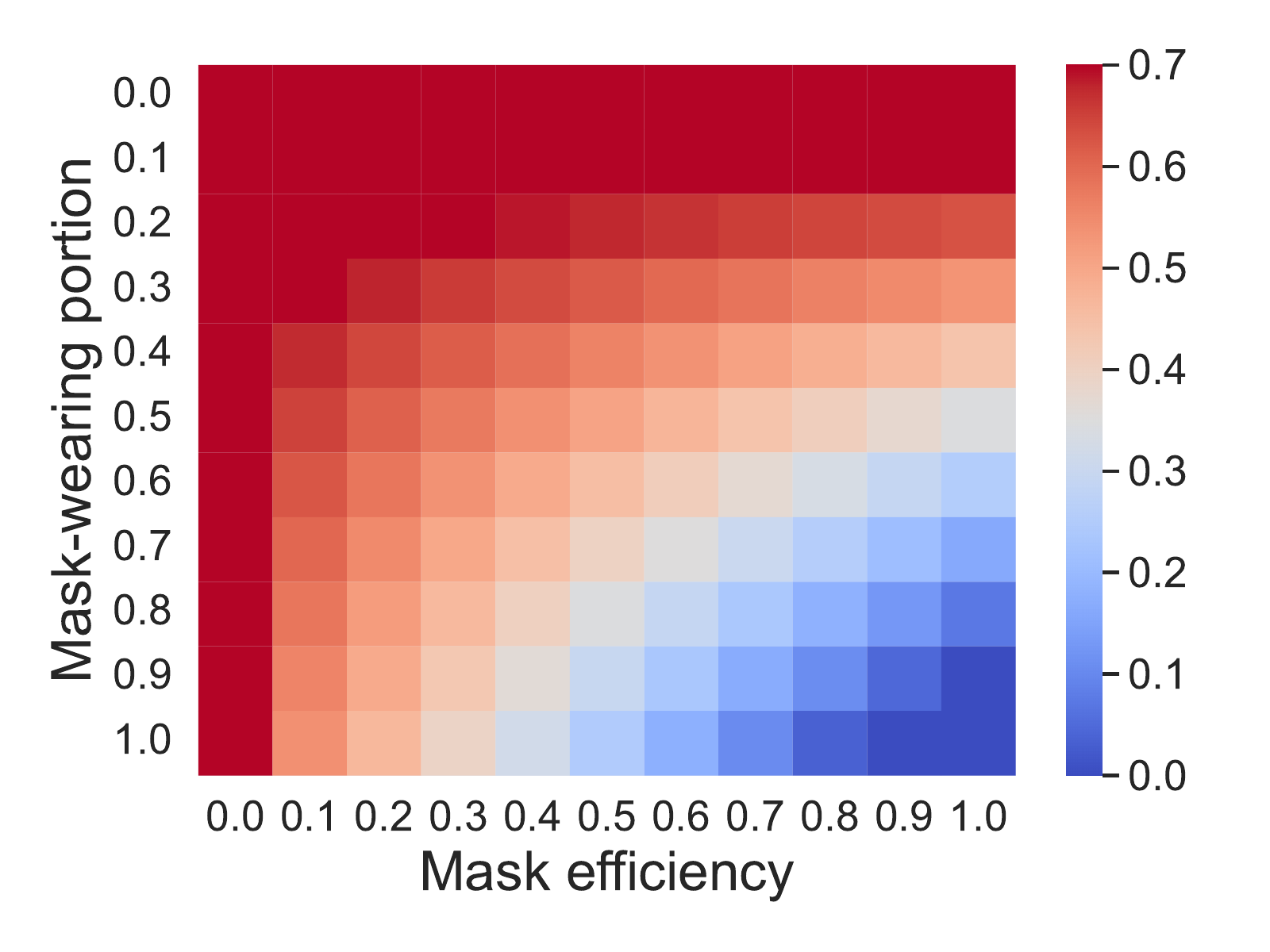}
    \caption{Conference - masks.}
    \label{fig:pip_conference_masks}
\end{subfigure}
\begin{subfigure}{.4\linewidth}
    \centering
    \includegraphics[width=0.75\linewidth]{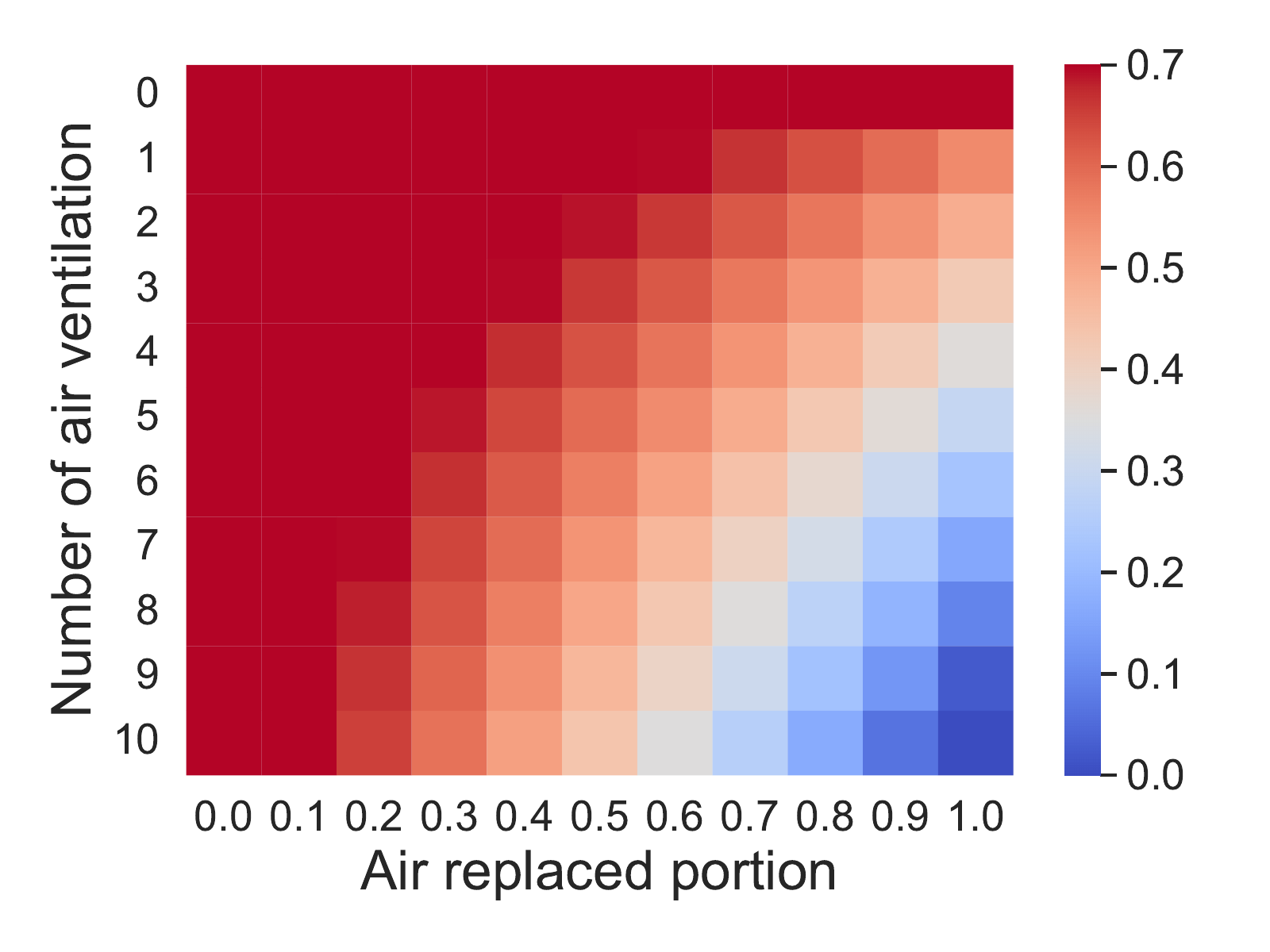}
    \caption{Conference - air ventilation.}
    \label{fig:pip_conference_air}
\end{subfigure}

\begin{subfigure}{.4\linewidth}
    \centering
    \includegraphics[width=0.75\linewidth]{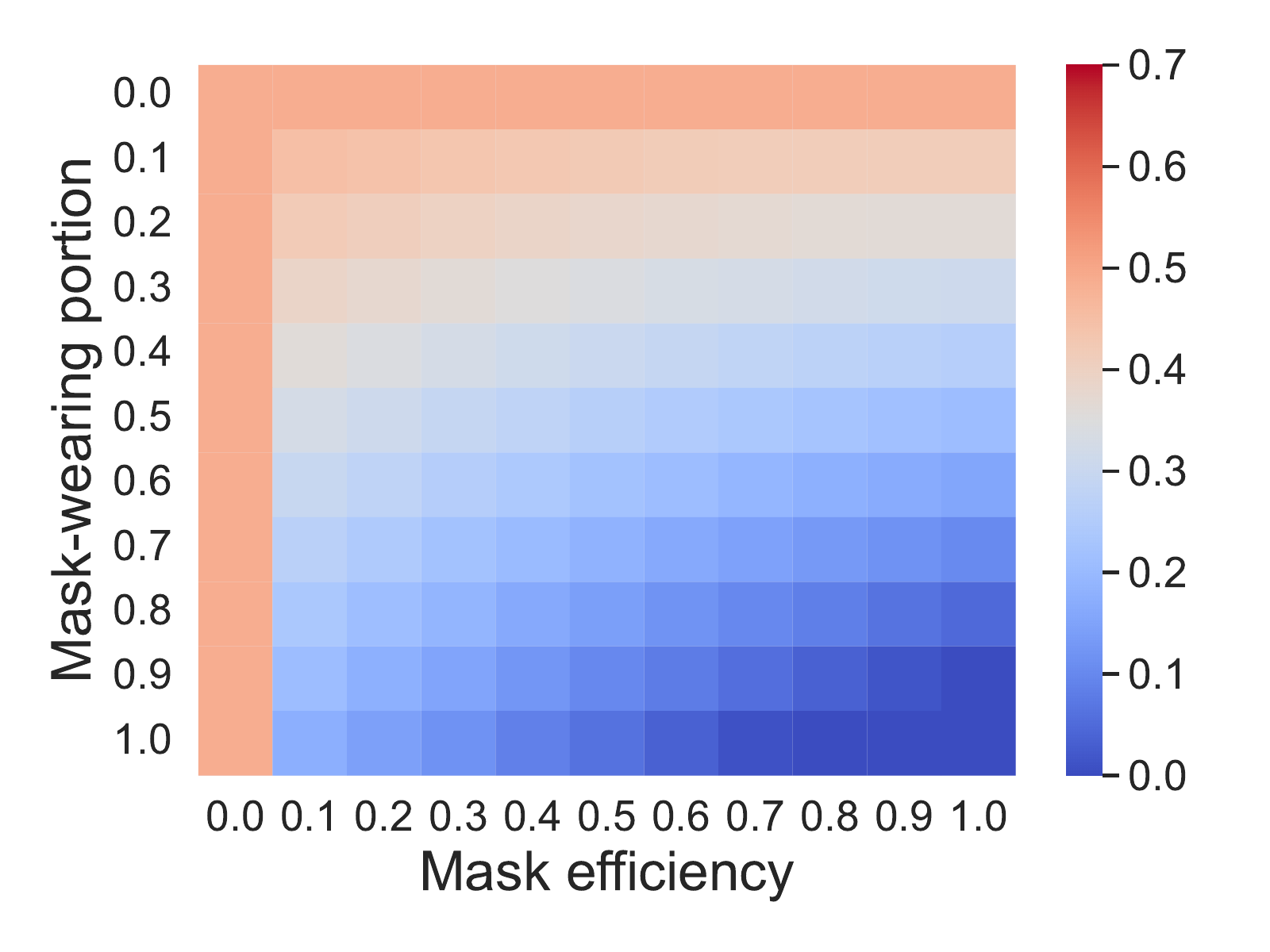}
    \caption{Movie theater - masks.}
    \label{fig:pip_movie_masks}
\end{subfigure}
\begin{subfigure}{.4\linewidth}
    \centering
    \includegraphics[width=0.75\linewidth]{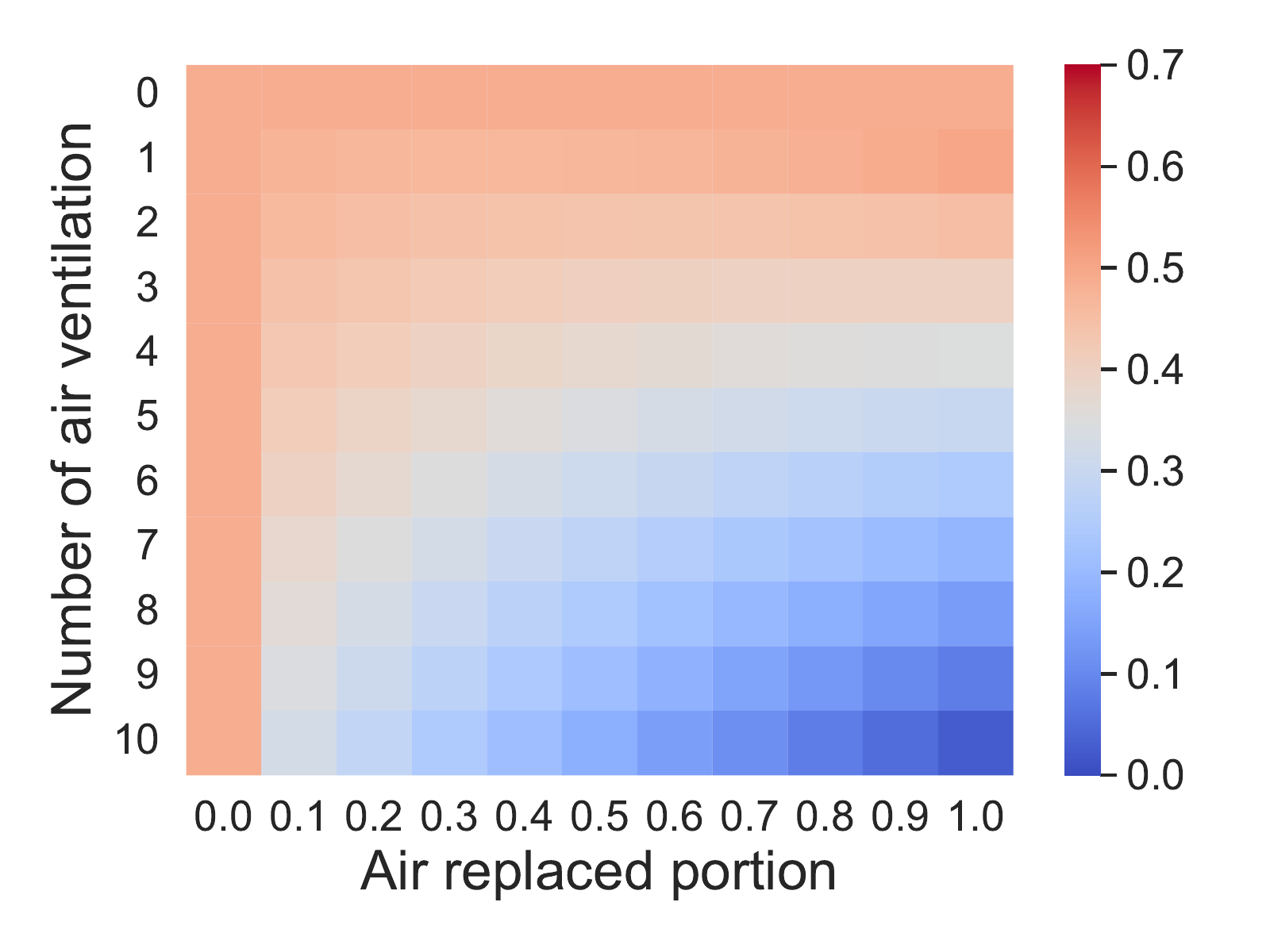}
    \caption{Movie theater - air ventilation.}
    \label{fig:pip_movie_air}
\end{subfigure}

\begin{subfigure}{.4\linewidth}
    \centering
    \includegraphics[width=0.75\linewidth]{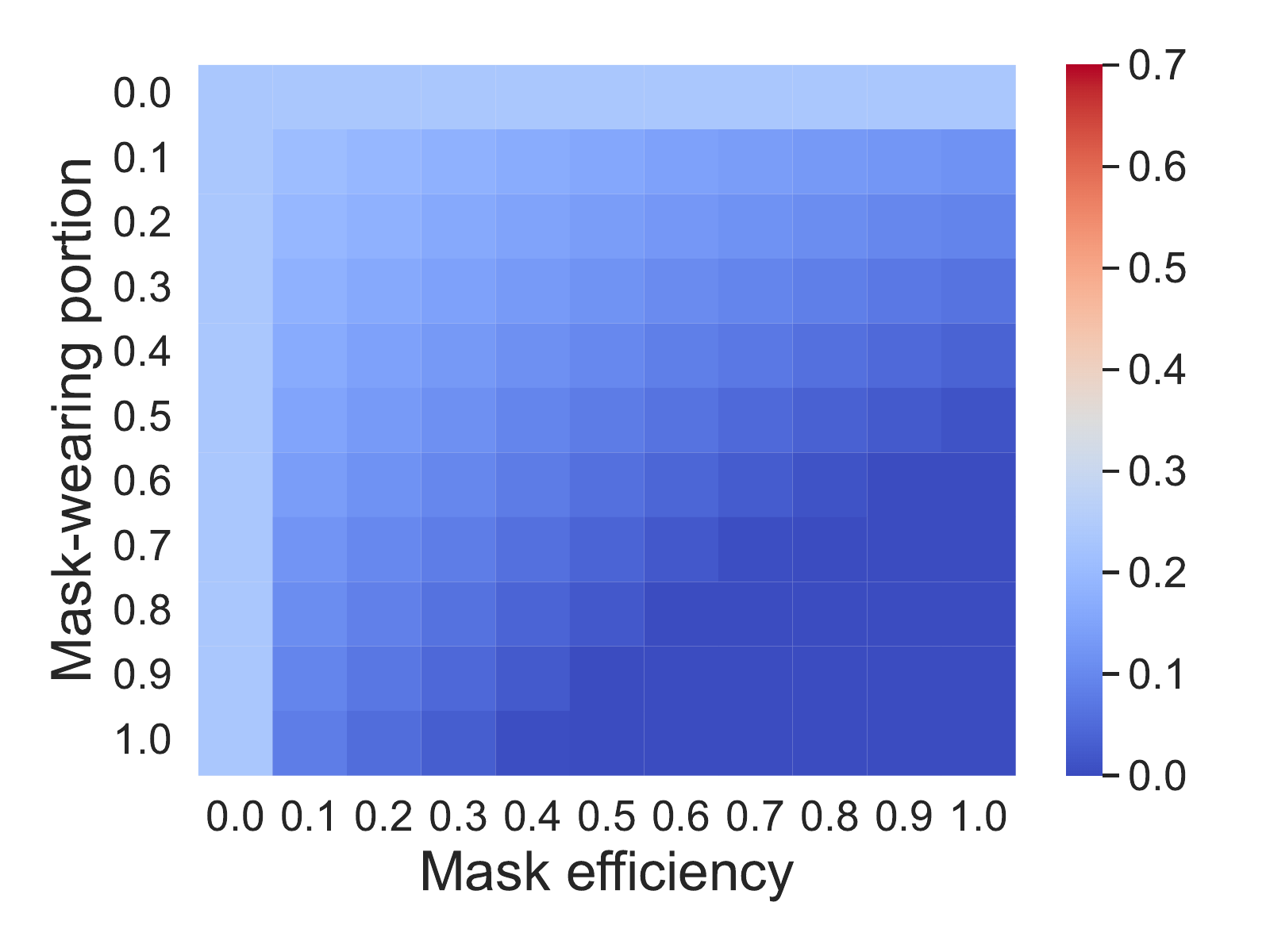}
    \caption{Restaurant - masks.}
    \label{fig:pip_restaurant_masks}
\end{subfigure}
\begin{subfigure}{.4\linewidth}
    \centering
    \includegraphics[width=0.75\linewidth]{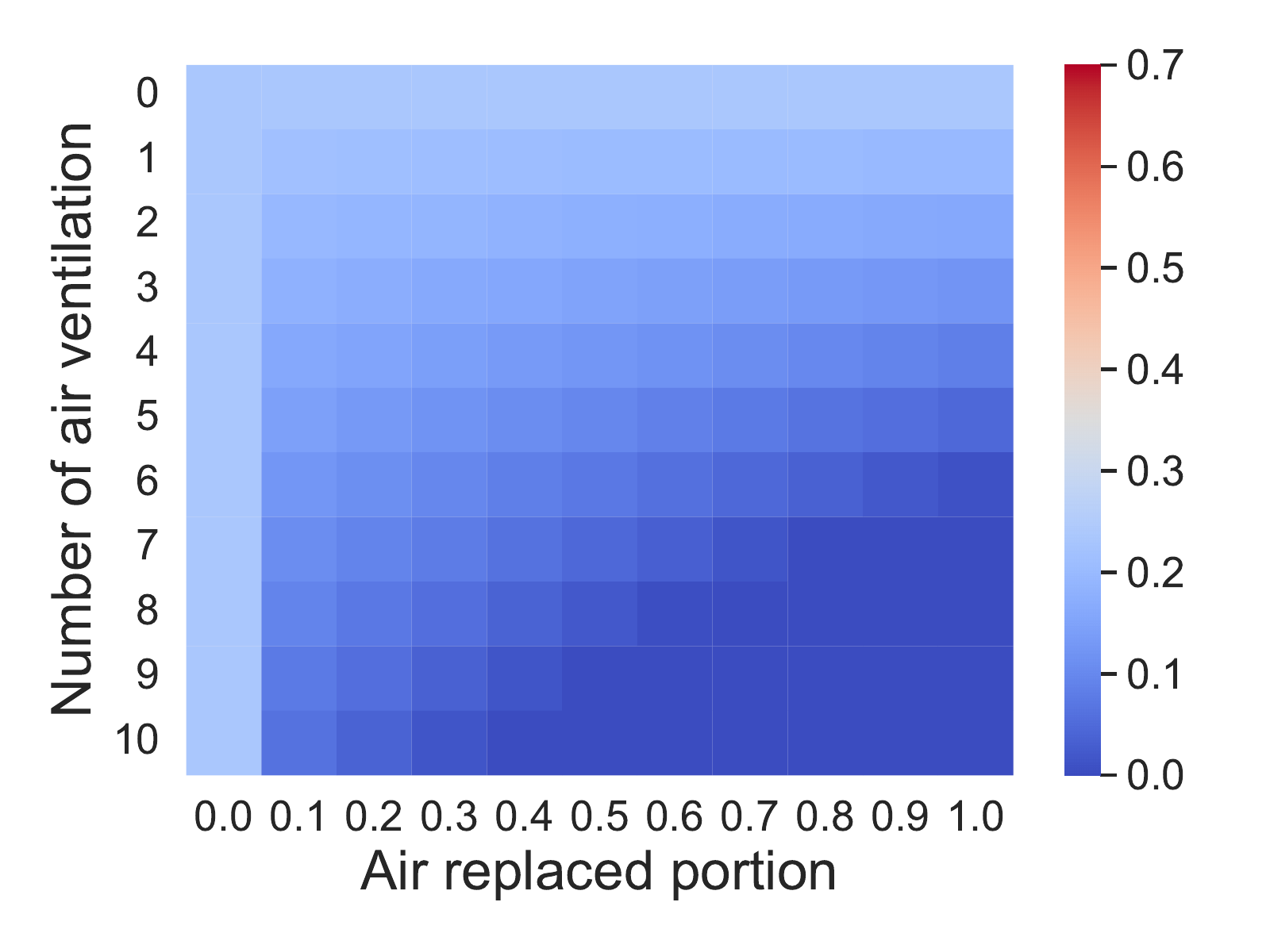}
    \caption{Restaurant - air ventilation.}
    \label{fig:pip_restaurant_air}
\end{subfigure}

\caption{The influence of the masks-wearing and air-ventilation PIPs on the number of infected individuals over a period of 90 minutes (an hour and a half). The results are shown as the average value of all the rooms of the same type.}
\label{fig:pip}
\end{figure}

Figs. \ref{fig:pip_classroom_masks}-\ref{fig:pip_restaurant_air} present the model sensitivity for both the mask-wearing and air-ventilation PIPs, divided into the four room types. For the mask-wearing PIP, the x-axis shows the rate of the average quality of mask-wearing in reducing infection rate (\(\eta\)) and the y-axis shows the portion of the population that wears masks (\(\kappa\)). Similarly, for the air-ventilation PIP, the x-axis shows the portion of the pathogen particles that are removed in each ventilation \(\chi\) and the y-axis shows the rate in minutes that the air-ventilation occurs \(\zeta\). The heatmaps present the average portion of exposed individuals out of the population after 90 minutes which is computed from 500 realisations (100 realisations for each room, five rooms for each room type). For each realisation, the infected individual is chosen randomly as well as the individuals that wear masks.

\section{Discussion}
\label{sec:discussion}
We have explored the airborne pandemic spread for a small size population, sharing a room using a combined CFD and \textit{SEI} model for the air flow and epidemiological dynamics, respectively. We evaluated the pandemic spread of four types of rooms: classroom, conference room, movie theater, and restaurant, using high-resolution spatial data obtained by a LiDAR and by simulating the air movement dynamics with a relatively small step in time to capture accurately all temporal dynamics.

We find that the spread dynamics between the four different room types are statistically different with p-value of \((0.0014\)) using Levene's test \cite{stat_1}. Moreover, we computed a pair-wise Welch's T-test showing that each pair of room types has statistically different dynamics with a worse p-value of \(0.046\) \cite{stat_2}. Thus, the pandemic spread dynamics have a consistent relationship with the distribution of the population in the room and its geometrical configuration, as shown in Fig. \ref{fig:baseline}. In particular, the population distribution has more influence on the pandemic spread rather than the overall density of the population, limited to scenarios similar to the ones examined in this study. For instance, the average restaurant's density is \(0.054\) [person/\(m^3\)] with \(0.033\) standard deviation with movie theaters' \(0.063 \pm 0.046\) [person/\(m^3\)] density, as computed from Table \ref{table:topologies}. Despite, the pandemic spread in both is statistically significantly different with a p-value of \(0.024\) (obtained from a two-sided paired T-test). Hence, not only the well-mixed assumption not holds for density-related approximations is not capturing the entire airborne pandemic spread dynamics even for a short duration as has been believed before \cite{density_1,density_2}. Nevertheless, for the case of a relatively large population size of a few hundred individuals found in the movie theater, our model predicts a similar pandemic spread rate as found by \cite{indoor_pandemic_model,review_1}. This can be explained by the influence of breathing zone infection versus indirect infection \cite{visualization_behavior_office_buildings}. One can see from Fig. \ref{fig:setup} that individuals in the conference room are facing each other and the intersection of the breathing zone is much larger compared to the movie theater where individuals are facing a screen and there is enough distance between any two individuals. As a result, as presented in Fig. \ref{fig:baseline}, the portion of exposed individuals in the conference room is much higher on average compared to the movie theater for the same duration. That said, these results provide an upper boundary to the realistic pandemic spread in restaurants and movie theaters as these have ventilation systems that reduce the infection rate.

When introducing mask-wearing or artificial air ventilation (AAV), we show that mask-wearing is consistently outperforming AAV in reducing the pandemic spread over all four room types, as shown in Fig. \ref{fig:pip}. Nonetheless, the mask-wearing PIP requires the participation of the population which is known to be harder to get over long periods of time \cite{masks_1,masks_2,masks_3}. On the other hand, even a low level of air ventilation like a replacement of 20\% of the air in the room every 30 minutes results in \(7.25\) percent improvement compared to a room without any air ventilation and can be controlled more easily compared to the mask-wearing. These results agree with the outcomes proposed by \cite{covid_with_ventilation}.

Our analysis shows that PIPs to limit shared room airborne pandemic spread are needed in most indoor spaces whenever COVID-19 is spreading in a community. These results are even more relevant with more contagious pathogens such as measles \cite{measles_1,measles_2}. The proposed model may be useful in the design and renovation of building systems. In particular, one can integrate it into graph-based spatio-temporal epidemiological models of buildings such as the one proposed by \cite{teddy_ariel} in order to obtain a more accurate pandemic spread dynamics. In this work, we study only the mask-wearing and AAV measures to reduce pandemic spread but other measures such as avoiding intense physical activities, shortening the duration of occupancy, social distance, and additional virus removal from the air devices should be taken into consideration when evaluating pandemic spread in a room-level and planning intervention policies. The combination of several PIPs are already taking place in practice and theory \cite{teddy_pandemic_management,multi_pip_1,multi_pip_2}, showing promising outcomes. The simultaneous usage of multiple PIPs in a single room is expected to have similar outcomes. 

Of note, the values reported in the analysis are based on the first (source) strain of the COVID-19 pandemic. Later strains of COVID-19 are known to be more aggressive with a higher infection rate (as reflected in the number of pathogen particles infectious individuals generate and the need for susceptible individuals to get infected) which might significantly alter the results. In addition, the computation of the CFD model is considered computationally expensive and therefore not feasible for large-scale models handling cities and countries. For this research, we used a server with four GTX 1080 Ti (Nvidia) GPUs and 16-Core (Intel Xeon) LGA 3647 CPU that computed for 147 hours to obtain the simulation outputs due to the high spatial and temporal resolution used in the simulations. Possible future research can study the trade-off between simulation accuracy and computational efficiency for the different spatial grid and time-step sizes. In addition, these results are obtained by the proposed model and simulator which are not validated on empirical data. While the CFD and \textit{SEI} models are validated separately multiple times in the past, the combination between them in general and in the context of breathing individuals in a room are not. That said, a  a complete validation of the proposed model is not feasible as one would need to populate a room with individuals and sample the number of infected individuals over time. However, an experiment that aims to infected individuals with a pathogen are unethical (for a good reason) and thus the experiment required to validate the model is impossible. One can argue that just simulating the airflow alone without the pandemic model can be sufficient. However, the novelty of the proposed model lies in the breathing dynamics and their influence on the infection rate which does not captured in such a validation deemed it to be insufficient. A possible walk-around of this issue is the usage of machine that mimics the breathing pattern of individuals. Nonetheless, the use of such a machine raise another validation issue where one needs to show that the proposed machine accurately reproduces human's breathing patterns. Moreover, as one would need to sample the dynamics multiple times (due to the stochastic behavior of individuals) with machine populations that includes several dozens to hundreds in order to reproduce the proposed results (see Table \ref{table:topologies}). This, by itself, would be extremely expensive and unreachable for most academic labs. 

\section{Conclusion}
\label{sec:conclusion}
The model developed in this study integrates a highly detailed 3D geometrical configuration of a room obtained using a LiDAR scanning device with a spatio-temporal pandemic spread model that is constructed from a \textit{SEIR}-based epidemiological model with a CFD-based air-flow model. The proposed model allows us to examine the influence of different room types on an airborne pandemic spread. In addition, the influence of masks-wearing and air ventilation in the room-level with relatively small populations and time duration on the pandemic spread is analyzed.

The proposed model is implemented for the COVID-19 outbreak. These spatio-temporal interactions allow one to explore the reciprocal effects of both spatial and temporal PIP on the spread of the pandemic in different social activities. An example is the effect of mask-wearing in the movie theater, as shown in Fig. \ref{fig:pip_movie_masks}. The inclusion of these interactions provides a highly detailed representation of the physical way airborne pathogens move from one host (individual) to another. Thus, these improve the accuracy of the model's forecasts and allow for multidimensional analysis of the impact of PIPs. 

Our results indicate that policymakers need to take into consideration the unique social properties that individuals in the population have in different rooms to obtain the optimal PIP. The optimal PIP for each room type provides policymakers with a differential mask-wearing and/or AAV intervention policy among different locations, such as in restaurants and classrooms. We find that the distribution of the population in the room influences AAV mainly by the amount of breathing zone infections. Thus, given a mask-wearing and AAV configuration, one can further control the pandemic spread by controlling breathing zone infection using social distance and other manners. 

The proposed model assumes that individuals are not moving, the AAV is affecting the entire air in the room at once, and the tempature is constant. In future work, one can relax these assumptions in order to obtain a more realistic representation of the airborne spread dynamics in a room. 

\section*{Declarations}
\subsection*{Funding}
This research did not receive any specific grant from funding agencies in the public, commercial, or not-for-profit sectors.

\subsection*{Data availability}
The room's raw topological data that has been used is not available due to security reasons, according to Israeli law. Besides that, the data that has been used is provided in the manuscript with the relevant sources.

\subsection*{Conflicts of interest}
The authors have no financial or proprietary interests in any material discussed in this article.

\subsection*{Ethics approval statement}
NA

\subsection*{Patient consent statement}
NA

\subsection*{Author Contributions}
Formal analysis and investigation, data gathering, and original draft preparation were performed by Ariel Alexi;
Conceptualization, formal analysis and investigation, coding, manuscript editing, and supervision were performed by Teddy Lazebnik.

\subsection*{Acknowledgments}
The authors wish to thank Noa Vardi for her statistical consulting. 

\bibliographystyle{unsrt}
\bibliography{bilbo.bib}

\begin{thebibliography}{10}

\bibitem{pandemic_duration}
A.~Brodeur, D.~Gray, A.~Islam, and S.~Bhuiyan.
\newblock A literature review of the economics of covid-19.
\newblock {\em IZA Discussion Paper No. 13411, Available at SSRN:
  https://ssrn.com/abstract=3636640}, 2020.

\bibitem{pandemic_important}
Andrea~Alberto Conti.
\newblock {Historical and methodological highlights of quarantine measures:
  from ancient plague epidemics to current coronavirus disease (COVID-19)
  pandemic}.
\newblock {\em Acta bio-medica : Atenei Parmensis}, 91(2):226--229, 2020.

\bibitem{who_problem}
{Eurosurveillance Editorial Team}.
\newblock Note from the editors: World health organization declares novel
  coronavirus (2019-ncov) sixth public health emergency of international
  concern.
\newblock {\em Euro Surveill}, 25:200131e, 2020.

\bibitem{who_data}
WHO.
\newblock Who coronavirus disease (covid-19) dashboard, 2022.

\bibitem{ravaging1988}
Joshua Lederberg.
\newblock {Medical Science, Infectious Disease, and the Unity of Humankind}.
\newblock {\em JAMA}, 260(5):684--685, 1988.

\bibitem{urbanization2017}
T.~Wu, C.~Perrings, A.~Kinzig, J.~P. Collins, B.~A. Minteer, and P.~Daszak.
\newblock Economic growth, urbanization, globalization, and the risks of
  emerging infectious diseases in china: A review.
\newblock {\em Ambio}, 46(1):18--29, 2017.

\bibitem{std_1}
T.~C. Quinn.
\newblock Global burden of the hiv pandemic.
\newblock {\em The Lancet}, 348(9020):99--106, 1996.

\bibitem{std_2}
S.~J. Genuis, F.~Dabog, and S.~K. Genuis.
\newblock Managing the sexually transmitted disease pandemic: A time for
  reevaluation.
\newblock {\em American Journal of Obstetrics and Gynecology},
  191(4):1103--1112, 2004.

\bibitem{social_pandemic}
S.~Djillali, S.~Bentout, T.~M. Touaoula, and A.~Tridane.
\newblock Global dynamics of alcoholism epidemic model with distributed delays.
\newblock {\em Mathematical Biosciences and Engineering}, 18, 2021.

\bibitem{social_pandemic_2}
S.~Djillali, S.~Bentout, T.~M. Touaoula, A.~Tridane, and S.~Kumar.
\newblock Global behavior of heroin epidemic model with time distributed delay
  and nonlinear incidence function.
\newblock {\em Results in Physics}, 31, 2021.

\bibitem{airborne_sample_1}
N.~M. Ferguson, D.~A.~T. Cummings, C.~Fraser, J.~C. Cajka, P.~C. Cooley, and
  D.~S. Burke.
\newblock Strategies for mitigating an influenza pandemic.
\newblock {\em Nature}, pages 448--452, 2006.

\bibitem{airborne_sample_2}
J.~L. Domingo, M.~Marques, and J.~Rovira.
\newblock Influence of airborne transmission of sars-cov-2 on covid-19
  pandemic. a review.
\newblock {\em Environmental Research}, 188:109861, 2020.

\bibitem{airborne_important}
J.~V. Fernandez-Montero, V.~Soriano, P.~Barreiro, C.~de~Mendoza, and M.~Á.
  Artacho.
\newblock Coronavirus and other airborne agents with pandemic potential.
\newblock {\em Current opinion in environmental science \& health}, 17:41--48,
  2020.

\bibitem{pip_1}
O.~M. Araz, P.~Damien, D.~A. Paltiel, S.~Burke, B.~van~de Geijn, A.~Galvani,
  and L.~A. MEyers.
\newblock Simulating school closure policies for cost effective pandemic
  decision making.
\newblock {\em BMC Public Health}, page 449, 2012.

\bibitem{pip_2}
M.~I. Meltzer, N.~J. Cox, and K.~Fukuda.
\newblock The economic impact of pandemic influenza in the united states:
  priorities for intervention.
\newblock {\em Emerging Infectious Diseases}, 5(5):659--671, 1999.

\bibitem{pip_3}
M.~Kabir, M.~S. Afzai, A.~Khan, and H.~Ahmed.
\newblock Covid-19 pandemic and economic cost; impact on forcibly displaced
  people.
\newblock {\em Travel Medicine and Infectious Disease}, 35:101661, 2020.

\bibitem{pip_4}
P.~Perrin, O.~McCabe, G.~Everly, and J.~Links.
\newblock Preparing for an influenza pandemic: Mental health considerations.
\newblock {\em Prehospital and Disaster Medicine}, 24(3), 2009.

\bibitem{pip_5}
M.~R. Taylor, K.~E. Agho, G.~J. Stevens, and B.~Raphael.
\newblock Factors influencing psychological distress during a disease epidemic:
  Data from australia's first outbreak of equine influenza.
\newblock {\em BMC Public Health}, 8:347, 2008.

\bibitem{models_good_2}
J.~C. Miller.
\newblock Mathematical models of sir disease spread with combined non-sexual
  and sexual transmission routes.
\newblock {\em Infectious Disease Modelling}, 2:35--55, 2017.

\bibitem{models_good_1}
A.~R. Tuite, D.~N. Fisman, and A.~L. Greer.
\newblock Mathematical modelling of covid-19 transmission and mitigation
  strategies in the population of ontario, canada.
\newblock {\em CMAJ}, 192:E497--E505, 2020.

\bibitem{first_sir}
W.~O. Kermack and A.~G. McKendrick.
\newblock A contribution to the mathematical theory of epidemics.
\newblock {\em Proceedings of the Royal Society}, 115:700–721, 1927.

\bibitem{first_teddy_paper}
T.~Lazebnik and S.~Bunimovich-Mendrazitsky.
\newblock The signature features of covid-19 pandemic in a hybrid mathematical
  model—implications for optimal work–school lockdown policy.
\newblock {\em Adv. Theory Simul.}, 4(5):e2000298, 2021.

\bibitem{different_approach_from_sir}
B.~Ivorra, M.~R. Ferrandez, M.~Vela-Perez, and A.~M. Ramos.
\newblock Mathematical modeling of the spread of the coronavirus disease 2019
  (covid-19) taking into account the undetected infections. the case of china.
\newblock {\em Commun Nonlinear Sci Numer Simulat}, 2020.

\bibitem{different_approach_from_sir_2}
Justin~B. Long and Jesse~M. Ehrenfeld.
\newblock The role of augmented intelligence (ai) in detecting and preventing
  the spread of novel coronavirus.
\newblock {\em Journal of Medical Systems}, 44, 2020.

\bibitem{different_approach_from_sir_3}
L.~Nesteruk.
\newblock Statistics-based predictions of coronavirus epidemic spreading in
  mainland china.
\newblock {\em Innov Biosyst Bioeng}, 4:13--18, 2020.

\bibitem{teddy_economic}
Teddy Lazebnik, Labib Shami, and Svetlana Bunimovich-Mendrazitsky.
\newblock Spatio-temporal influence of non-pharmaceutical interventions
  policies on pandemic dynamics and the economy: The case of covid-19.
\newblock {\em Research Economics}, 2021.

\bibitem{sir_clinical}
D.~Acemoglu, V.~Chernozhukov, I.~Werning, and M.~D. Whinston.
\newblock Optimal targeted lockdowns in a multigroup sir model.
\newblock {\em American Economic Review: Insights}, 3(4):487--502, 2021.

\bibitem{teddy_ariel}
T.~Lazebnik and A.~Alexi.
\newblock Comparison of pandemic intervention policies in several building
  types using heterogeneous population model.
\newblock {\em Communications in Nonlinear Science and Numerical Simulation},
  107(4):106176, 2022.

\bibitem{graph_2}
W.~J. Edmunds, C.~J. O'Callaghan, and D.~J. Nokes.
\newblock Who mixes with whom? a method to determine the contact patterns of
  adults that may lead to the spread of airborne infections.
\newblock {\em Proceedings. Biological sciences}, 264(1384):949--957, 1997.

\bibitem{spatial_1}
M.~J. Keeling.
\newblock The implications of network structure for epidemic dynamics.
\newblock {\em Theoretical Population Biology}, 67(1):1--8, 2005.

\bibitem{graph_4}
A.~S. Klovdahl, J.~J. Potterat, D.~E. Woodhouse, J.~B. Muth, S.~Q. Muth, and
  W.~W. Darrow.
\newblock Social networks and infectious disease: The colorado springs study.
\newblock {\em Social Science \& Medicine}, 38(1):79--88, 1994.

\bibitem{spatial_2}
M.~J. Keeling and K.~T.~D. Eames.
\newblock Networks and epidemic models.
\newblock {\em Journal of The Royal Society Interface}, 2:295--307, 2005.

\bibitem{spatial_3}
P.~S. Bearman, J.~Moody, and K.~Stovel.
\newblock Chains of affection: the structure of adolescent romantic and sexual
  networks.
\newblock {\em Am. J. Sociol.}, 110:44--91, 2004.

\bibitem{cooper}
I.~Cooper, A.~Mondal, and C.~G. Antonopoulos.
\newblock A {SIR} model assumption for the spread of covid-19 in different
  communities.
\newblock {\em Chaos, Solitons \& Fractals}, 139:110057, 2020.

\bibitem{teddy_pandemic_management}
T.~Lazebnik, S.~Bunimovich-Mendrazitsky, and L.~Shami.
\newblock Pandemic management by a spatio–temporal mathematical model.
\newblock {\em International Journal of Nonlinear Sciences and Numerical
  Simulation}, 107(4):106176, 2021.

\bibitem{graph_3}
C.~Moore and M.~E.~J. Newman.
\newblock Epidemics and percolation in small-world networks.
\newblock {\em arXiv}, 2000.

\bibitem{spatial_example_1}
F.~A. Milner and R.~Zhao.
\newblock S-i-r model with directed spatial diffusion.
\newblock {\em Mathematical Population Studies}, 15(3), 2008.

\bibitem{spatial_example_2}
G.~Fabricius and A.~Maltz.
\newblock Exploring the threshold of epidemic spreading for a stochastic sir
  model with local and global contacts.
\newblock {\em Physica A: Statistical Mechanics and its Applications},
  540:123208, 2020.

\bibitem{spatial_example_3}
H.~Paeng, S and J.~Lee.
\newblock Continuous and discrete sir-models with spatial distributions.
\newblock {\em J. Math. Biol.}, 74:1709--1727, 2017.

\bibitem{airflow_pandemic}
S.~Samaresh, V.~Vaishali, V.~R. Abhay, S.~Aditya, K.~Abhishek, and V.~A. Shail.
\newblock Real-time imaging of airflow patterns and impact of infection control
  measures in ophthalmic practice: a pandemic perspective.
\newblock {\em Journal of Cataract \& Refractive Surgery}, 47(7):842--846,
  2021.

\bibitem{indoor_pandemic}
J.~Wei and Y.~Li.
\newblock Airborne spread of infectious agents in the indoor environment.
\newblock {\em American Journal of Infection Control}, 44(6):S102--S108, 2016.

\bibitem{airflow_health}
R.~A. Segal, X.~Guan, M.~Shearer, and T.~B. Martonen.
\newblock Mathematical model of airflow in the lungs of children i; effects of
  tumor sizes and locations.
\newblock {\em Journal of Theoretical Medicine}, 2(3):199--213, 2000.

\bibitem{airflow_mechanics}
H.D. Ammari.
\newblock A mathematical model of thermal performance of a solar air heater
  with slats.
\newblock {\em Renewable Energy}, 28(10):1597--1615, 2003.

\bibitem{airflow_agriclature}
C.~Rossello, J.~Canellas, S.~Simal, and A.~Berna.
\newblock Simple mathematical model to predict the drying rates of potatoes.
\newblock {\em J. Agrie. Food Chem.}, 40:2374--2378, 1992.

\bibitem{indoor_pandemic_model}
Z.~Peng, P.~A.~L. Rogas, E.~Kropff, W.~Bahnfleth, G.~Buonanno, S.~J. Dancer,
  J.~Kurnitski, Y.~Li, M.~G. L.~C. Loomans, L.~C. Marr, L.~Morawska,
  W.~Nazaroff, C.~Noakes, X.~Querol, C.~Sekhar, R.~Tellier, T.~Greenhalgh,
  L.~Bourouiba, A.~Boerstra, J.~W. Tang, S.~L. Miller, and J.~L. Jimenez.
\newblock Practical indicators for risk of airborne transmission in shared
  indoor environments and their application to covid-19 outbreaks.
\newblock {\em Environmental Science and Technology}, 56:1125--1137, 2020.

\bibitem{wells_riley}
E.~C. Riley, G.~Murphy, and R.~L. Riley.
\newblock Airborne spread of measles in a suburban elementary school.
\newblock {\em American Journal of Epidemiology}, 107:421--432, 1978.

\bibitem{airflow_building_review}
Y.~Yu, A.~C. Megri, and S.~Jiang.
\newblock A review of the development of airflow models used in building load
  calculation and energy simulation.
\newblock {\em Building Simulation}, 12(7):347–363, 2019.

\bibitem{airflow_building_review_2}
H.~Fariborz and H.~Li.
\newblock Building airflow movement - validation of three airflow models.
\newblock {\em Journal of Architectural and Planning Research},
  21(4):331–349, 2004.

\bibitem{ns_3d}
C.~Cao.
\newblock Sufficient conditions for the regularity to the 3d navier-stokes.
\newblock {\em Discrete and Continuous Dynamical System}, 26(4):1141–1151,
  2016.

\bibitem{cfd_intro}
T.~A.~G. Smyth.
\newblock A review of computational fluid dynamics (cfd) airflow modelling over
  aeolian landforms.
\newblock {\em Aeolian Research}, 22:153--164, 2016.

\bibitem{airflow_building_models}
H.~Versteeg and W.~Malalasekra.
\newblock An introduction to computational fluid dynamics — the finite volume
  method.
\newblock {\em New York: Pearson Education}, 2007.

\bibitem{rans}
F.~Su, S.~A. Kinnas, and H.~Jukola.
\newblock Application of a bem/rans interactive method to contra-rotating
  propellers.
\newblock {\em Fifth International Symposium on Marine Propulsion}, 2017.

\bibitem{cfd_summary}
Y.~Zhiyin.
\newblock Large-eddy simulation: Past, present and the future.
\newblock {\em Chinese Journal of Aeronautics}, 28(1):11--24, 2015.

\bibitem{cfd_1}
S.~Kato.
\newblock Review of airflow and transport analysis in building using cfd and
  network model.
\newblock {\em Japan Architectural Review}, 1:299--309, 2018.

\bibitem{cfd_2}
H.~B. Nahor, M.~L. Hoang, P.~Verboven, M.~Baelmans, and B.~M. Nicolai.
\newblock Cfd model of the airflow, heat and mass transfer in cool stores.
\newblock {\em International Journal of Refrigeration}, 28:368--380, 2005.

\bibitem{cfd_3}
N.~J. Smale, J.~Moureh, and G.~Cortella.
\newblock A review of numerical models of airflow in refrigerated food
  applications.
\newblock {\em International Journal of Refrigeration}, 29:911--930, 2006.

\bibitem{cfd_cool}
C.~Cravero and D.~Marsano.
\newblock Simulation of covid-19 indoor emissions from coughing and breathing
  with air conditioning and mask protection effects.
\newblock {\em Indoor and Built Environment}, 2021.

\bibitem{ns}
C.~L. Fefferman.
\newblock Existence and smoothness of the navier-stokes equation.
\newblock {\em Computers \& Fluids}, 7:86--92, 2013.

\bibitem{particale_decay}
J.~Zheng, X.~Wu, F.~Fang, J.~Li, Z.~Wang, H.~Xiao, J.~Zhu, C.~Pain, P.~Linden,
  and B.~Xiang.
\newblock Numerical study of covid-19 spatial–temporal spreading in london.
\newblock 33:E10, 2021.

\bibitem{particale_decay_2}
A.~W.~H. Chin, J.~T.~S. Chu, M.~R.~A. Perera, K.~P.~Y. Hui, H-L. Yen, M.~C.~W.
  Chan, M.~Peiris, and L.~L.~M. Poon.
\newblock Stability of sars-cov-2 in different environmental conditions.
\newblock 1(1):E10, 2020.

\bibitem{agent_based_with_ai}
G.~Ciatto, M.~I. Schumacher, A.~Omicini, and D.~Calvaresi.
\newblock Agent-based explanations in {AI}: Towards an abstract framework.
\newblock In {\em International Workshop on Explainable, Transparent Autonomous
  Agents and Multi-Agent Systems}, pages 3--20. Springer, 2020.

\bibitem{agent_based_exp_1}
L.~Tesfatsion.
\newblock Agent-based computational economics: Growing economies from the
  bottom up.
\newblock {\em Artificial Life}, 8(1), 2002.

\bibitem{agent_simulation_1}
M.~Raberto, S.~Cincotti, S.~M. Focardi, and M.~Marchesi.
\newblock Agent-based simulation of a financial market.
\newblock {\em Physica A: Statistical Mechanics and its Applications},
  299:319--327, 2001.

\bibitem{human_behavior_in_buildings}
C.~Peng, D.~Yan, R.~Wu, C.~Wang, X.~Zhou, and Y.~Jiang.
\newblock Quantitative description and simulation of human behavior in
  residential buildings.
\newblock 5:85–94, 2012.

\bibitem{restaurant}
K.~S. Kwon, J.~I. Park, Y.~J. Park, D.~M. Jung, K.~W. Ryu, and J.~H. Lee.
\newblock Evidence of long-distance droplet transmission of sars-cov-2 by
  direct air flow in a restaurant in korea.
\newblock {\em J Korean Med Sci}, 35(46):e415, 2020.

\bibitem{restaurant_2}
Z.~Peng and J.~L. Jimenez.
\newblock Exhaled co2 as a covid-19 infection risk proxy for different indoor
  environments and activities.
\newblock {\em Environ. Sci. Technol. Lett.}, 8(5):392–397, 2021.

\bibitem{other_buildings_1}
Y.~Shen, C.~Li, H.~Dong, Z.~Wang, L.~Martinez, Z.~Sun, A.~Handel, Z.~Chen,
  E.~Chen, M.~H. Ebell, F.~Wang, B.~Yi, H.~Wang, X.~Wang, A.~Wang, B.~Chen,
  Y.~Qi, L.~Liang, Y.~Li, F.~Ling, J.~Chen, and G.~Xu.
\newblock Community outbreak investigation of sars-cov-2 transmission among bus
  riders in eastern china.
\newblock {\em JAMA Internal Medicine}, 180(12):1665--1671, 2020.

\bibitem{other_buildings_2}
S.~Jang, S.~Han, and J.~Rhee.
\newblock Cluster of coronavirus disease associated with fitness dance classes,
  south korea.
\newblock {\em Emerging Infectious Diseases}, 26(8):1917--1920, 2020.

\bibitem{lidar}
S.~Jie and A.~Sampath.
\newblock Urban dem generation from raw lidar data.
\newblock {\em Photogrammetric Engineering \& Remote Sensing}, pages 217--226,
  2005.

\bibitem{point_to_mesh}
B.~Douillard, J.~Underwood, N.~Kuntz, V.~Vlaskine, A.~Quadros, P.~Morton, and
  A.~Frenkel.
\newblock On the segmentation of 3d lidar point clouds.
\newblock In {\em 2011 IEEE International Conference on Robotics and
  Automation}, pages 2798--2805, 2011.

\bibitem{particale_decay_3}
J.~Zheng, J.~Zhu, Z.~Wang, F.~Fang, C.~C. Pain, and J.~Xiang.
\newblock Towards a new multiscale air quality transport model using the fully
  unstructured anisotropic adaptive mesh technology of fluidity.
\newblock 8(10):3421--3440, 2015.

\bibitem{breathing_time}
J.~Hutchinson.
\newblock Breathing pattern in humans: diversity and individuality.
\newblock 1850.

\bibitem{breathing_volume}
W.~Perez and M.~J. Tobim.
\newblock Separation of factors responsible for change in breathing pattern
  induced by instrumentation.
\newblock 59:1515--1520, 1985.

\bibitem{breathing_time_2}
M.~A. Quetelet.
\newblock A treatise on man and the development of his faculties.
\newblock 1842.

\bibitem{breathing}
G.~Benchetrit.
\newblock Breathing pattern in humans: diversity and individuality.
\newblock 122(2-3):123--129, 2000.

\bibitem{bio_data}
E.~A. Hernandez-Vargas and J.~X. Velasco-Hernandez1.
\newblock In-host modelling of covid-19 kinetics in humans.
\newblock {\em medRxiv}, 2021.

\bibitem{covid_talbe_data}
R.~Sender, Y.~M. Bar-On, S.~Gleizer, B.~Bernshtein, A.~Flamholz, R.~Phillips,
  and R.~Milo.
\newblock The total number and mass of sars-cov-2 virions.
\newblock {\em Proceedings of the National Academy of Sciences},
  118(25):e2024815118, 2021.

\bibitem{lockdown_sir}
O.~Aglar, A.~Baxter, P.~Keskinocak, J.~Asplund, and N.~Serban.
\newblock Homebound by covid19: The benefits and consequences of
  non-pharmaceutical intervention strategies.
\newblock {\em Research Square}, 2020.

\bibitem{masks_good}
T.~Li, Y.~Liu, M.~Li, X.~Qian, and S.~Y. Dai.
\newblock Mask or no mask for covid-19: A public health and market study.
\newblock {\em PLoS ONE}, 15:e0237691, 2020.

\bibitem{influenza_mask}
N.~C.~J. Brienen, A.~Timen, J.~Wallinga, J.~E. Van~Steenbergen, and P.~F.~M.
  Teunis.
\newblock The effect of mask use on the spread of influenza during a pandemic.
\newblock {\em Risk Analysis}, 30(8):1210--1218, 2010.

\bibitem{close_lockdown}
Buse~Eylul Oruc, Arden Baxter, Pinar Keskinocak, John Asplund, and Nicoleta
  Serban.
\newblock Homebound by covid19: The benets and consequences of
  non-pharmaceutical intervention strategies.
\newblock {\em Research Square}, 2020.

\bibitem{masks_n95}
K.~O'Dowd, K.~M. Nair, P.~Forouzanadeh, S.~Mathew, J.~Grant, R.~Moran,
  J.~Bartlett, J.~Bird, and S.~C. Pillai.
\newblock Face masks and respirators in the fight against the covid-19
  pandemic: A review of current materials, advances and future perspectives.
\newblock {\em Materials}, 13:3363, 2020.

\bibitem{stat_1}
W-Y. Loh.
\newblock Some modifications of levene's test of variance homogeneity.
\newblock {\em Journal of Statistical Computation and Simulation},
  28(3):213--226, 1987.

\bibitem{stat_2}
R.~M. West.
\newblock Best practice in statistics: Use the welch t-test when testing the
  difference between two groups.
\newblock {\em Annals of Clinical Biochemistry}, 58(4):267--269, 2021.

\bibitem{density_1}
A.~R. Khavarian-Garmsir, A.~Sharifi, and N.~Moradpour.
\newblock Are high-density districts more vulnerable to the covid-19 pandemic?
\newblock {\em Sustainable Cities and Society}, 70:102911, 2021.

\bibitem{density_2}
S.~Hamidi, S.~Sabouri, and R.~Ewing.
\newblock Does density aggravate the covid-19 pandemic? early findings and
  lessons for planners.
\newblock {\em Journal of the American Planning Association}, 84(4):495--509,
  2020.

\bibitem{review_1}
T.~Fukuoka and K.~Ito.
\newblock Exposure risk assessment by coupled analysis of cfd and sir model in
  enclosed space.
\newblock 2010.

\bibitem{visualization_behavior_office_buildings}
Y.~Chen, X.~Liang, T.~Hong, and X.~Luo.
\newblock Simulation and visualization of energy-related occupant behavior in
  office buildings.
\newblock {\em Building Simulation}, 10:785–798, 2017.

\bibitem{masks_1}
M.~O. Rieger.
\newblock To wear or not to wear? factors influencing wearing face masks in
  germany during the covid-19 pandemic.
\newblock {\em Social Health and Behavior}, 3(2):50--54, 2020.

\bibitem{masks_2}
J.~H. Flaskrud.
\newblock Masks, politics, culture and health.
\newblock {\em Issues in Mental Health Nursing}, 41(9):846--849, 2020.

\bibitem{masks_3}
L.~Martinelli, V.~Kopilas, M.~Vidmar, C.~Heavin, H.~Machado, Z.~Todorovic,
  N.~Buzas, M.~Pot, B.~Prainsack, and S.~Gajovic.
\newblock Face masks during the covid-19 pandemic: A simple protection tool
  with many meanings.
\newblock {\em Frontiers in Public Health}, 8(9):846--849, 2021.

\bibitem{covid_with_ventilation}
H.~Dai and B.~Zhao.
\newblock Association of the infection probability of covid-19 with ventilation
  rates in confined spaces.
\newblock {\em Building Simulation}, 13:1321–1327, 2020.

\bibitem{measles_1}
H.~Levine and C.~Stein-Zamir.
\newblock The measles outbreak in israel in 2018-19: lessons for covid-19
  pandemic.
\newblock {\em Human Vaccines \& Immunotherapeutics}, 7:2085--2089, 2021.

\bibitem{measles_2}
L.~Roberts.
\newblock Why measles deaths are surging -- and coronavirus could make it
  worse.
\newblock {\em Nature}, 580(7804):446--447, 2020.

\bibitem{multi_pip_1}
N.~M. Ferguson, D.~Laydon, G.~Nedjati-Gilani, N.~Imai, K.~Ainslie, M.~Baguelin,
  S.~Bhatia, A.~Boonyasiri, Z.~Cucunuba, G.~Cuomo-Dannenburg, A.~Dighe,
  I.~Dorigatti, H.~Fu, K.~Gaythorpe, W.~Green, A.~Hamlet, W.~Hinsley, L.~C.
  Okell, S.~van Elsland, H.~Thompson, R.~Verity, E.~Volz, H.~Wang, Y.~Wang,
  P.~G.~T. Walker, C.~Walters, P.~Winskill, C.~Whittaker, C.~A. Donnelly,
  S.~Riley, and A.~C. Ghani.
\newblock Impact of non-pharmaceutical interventions (npis) to reduce covid-19
  mortality and healthcare demand.
\newblock {\em Imperial College}, 2020.

\bibitem{multi_pip_2}
J.~C. Lemaitre, K.~H. Grantz, J.~Kaminsky, H.~R. Meredith, S.~A. Truelove,
  S.~A. Lauer, L.~T. Keegan, S.~Shah, J.~Wills, K.~Kaminsky, J.~Perez-Saez,
  J.~Lessler, and E.~C. Lee.
\newblock A scenario modeling pipeline for covid-19 emergency planning.
\newblock {\em Scientific Reports}, 11:7534, 2021.

\end{thebibliography}

\end{document}